\documentclass[twocolumn, 10pt, aps, prl, superscriptaddress]{revtex4-1}

\usepackage{graphicx}
\usepackage{amsmath}
\usepackage{amssymb}
\usepackage[usenames, dvipsnames]{xcolor}
\usepackage{appendix}
\usepackage[colorlinks=true, pdfstartview=FitV, linkcolor=blue, citecolor=blue, urlcolor=blue]{hyperref}
\usepackage{color}
\usepackage{epstopdf}
\usepackage{dsfont}

\usepackage[bbgreekl,cspex]{mathbbol}
\usepackage{amsfonts}
\DeclareSymbolFontAlphabet{\mathbb}{AMSb}
\DeclareSymbolFontAlphabet{\mathbbl}{bbold}

\renewcommand{\Im}{\operatorname{Im}}
\DeclareMathOperator{\tr}{tr}
\DeclareMathOperator{\sgn}{sgn}

\begin{document}

\title{	
	Decoding the drive-bath interplay: A guideline to enhance superconductivity
}
 
\author{Rui Lin}
\affiliation{Institute for Theoretical Physics, ETH Z\"urich, 8093 Zurich, Switzerland}
\author{Aline Ramires}
\affiliation{Paul Scherrer Institut, 5232 Villigen PSI, Switzerland}
\author{R. Chitra}
\affiliation{Institute for Theoretical Physics, ETH Z\"urich, 8093 Zurich, Switzerland}

\date{\today}

 
\begin{abstract}
Driven-dissipative physics lie at the core of quantum optics. However, the full interplay between a driven quantum many-body system and its environment remains relatively unexplored in the solid state realm. In this Letter, we inspect this interplay beyond the commonly employed stroboscopic Hamiltonian picture based on the specific example of a driven superconductor. Using the Shirley-Floquet and Keldysh formalisms as well as a generalization of the notion of superconducting fitness to the driven case, we show how a drive which anti-commutes with the superconducting gap operator generically induces an unusual particle-hole structure in the spectral functions from the perspective of the thermal bath.  Concomitant with a driving frequency which is near resonant with the intrinsic cutoff frequency of the underlying interaction, this spectral structure can be harnessed to enhance the superconducting transition temperature. Our work paves the way for further studies for driven-dissipative engineering of exotic phases of matter in solid-state systems.
\end{abstract}
 
\maketitle

In the past decade, controllable light-matter coupling and  Floquet engineering have emerged as  powerful tools to tailor a plethora of  phenomena in quantum many-body systems.  This has permitted the exploration of novel out-of-equilibrium physics in the realm of quantum simulation as well as solid-state platforms. In the context of superconductivity, these tools  have been shown to  enhance~\cite{komnik16,knap16,sentef16,coulthard17,ido17} or induce superconductivity~\cite{mitrano16,kennes19,buzzi20,budden21}, and even generate exotic orders. These include non-trivial topology~\cite{takasan17,zhang21,dehghani21,kitamura22}, odd-frequency correlations~\cite{triola16,cayao21}, $\eta$-pairing~\cite{kaneko19,tindall19,ejima20,murakami22}, entropy-cooling mechanism~\cite{werner19}, as well as Amp\`{e}rean~\cite{schlawin19} and chiral superconductivity~\cite{li23}.  Most of these Floquet engineering schemes rely on the effective stroboscopic Hamiltonians~\cite{floquet83,shirley65,ivanov21}, which are renormalized, acquire new terms, or obtain an extra synthetic dimension~\cite{ozawa19,price15,price17,lustig19}.

Meanwhile, dissipative environments are ubiquitous, and help mitigate the problem of heating endemic to most interacting driven systems. Dissipation has also been developed as a resource  to engineer correlated steady states  especially for quantum computation applications~\cite{verstraete09}.
Combining both drive and  thermal dissipation paves the way for the exploration of surprising phenomena associated with the {\it relative rotation} between the system and the bath, which is intrinsically beyond the scope of effective stroboscopic Hamiltonians. In quantum optical-gaseous systems, dissipation is well captured by the rotating wave approximation and thus the Lindblad formalism, as the typical system energy scales (${<\mathrm{GHz}}$) are much smaller than the driving frequency (${\sim\mathrm{THz}}$)~\cite{sieberer16,manzano20}. This formalism predicts  unexpected driven-dissipative effects such as dissipative freezing~\cite{sanchez19}, quantum synchronization~\cite{tindall20},  modified critical behaviors~\cite{nagy11,brennecke13,torre13} and a stability towards high-energy steady states~\cite{soriente18,soriente20,ferri21,lin22,rosamedina22}.

 \begin{figure}[t]
	\centering
	\includegraphics[width=\columnwidth]{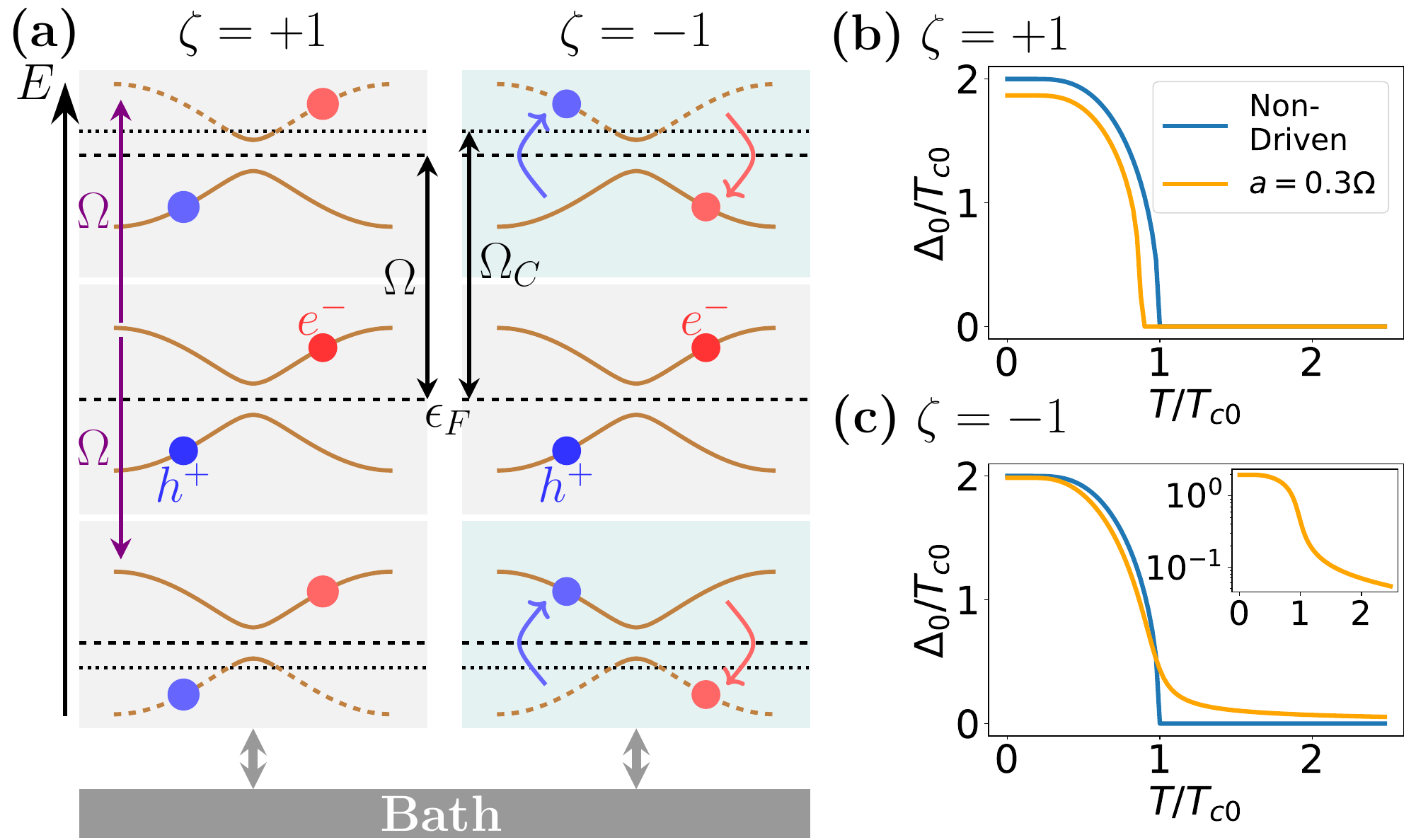}
	\caption{
			(a) Schematic representations of the Floquet band structure of the driven superconductor, when the drive (left) commutes and (right) anti-commutes with the gap, cf.\ Eq.~\eqref{eq:commutativity}. In the anti-commuting case, the roles of particles and holes are exchanged in every alternating temporal Brillouin zone. This ensures the excitations unfavorable to superconductivity lie outside of the cutoff frequency $\Omega_C$, and consequently, the system is less susceptible to temperature change of the bath.
			(b,c) The time-averaged gap $\Delta_0$ as a function of temperature $T$ for (b) $\zeta=+1$ ($H_1=\frac{a}{2}\tau_0\sigma_y$) and (c) $\zeta=-1$ ($H_1=\frac{a}{2}\mathrm{sgn}(k_x)\tau_z\sigma_y$). In panel (c), $\Delta_0$ is also plotted in logarithmic scale in the inset.
			Here, Eq.~\eqref{eq:gap_equation} is solved for flat-band superconductors Eq.~\eqref{eq:flat_band_hamil}.
	}
	\label{fig:schematics}
\end{figure}

In contrast, the intrinsic energy scales of solid-state systems (${\sim100\,\mathrm{GHz}}$) are comparable to the terahertz  driving frequencies in state-of-the-art experiments, making the Lindbladian approach insufficient. A more general description of the complex interplay between the drive and the thermal bath requires a combination of the Floquet~\cite{floquet83,shirley65} and Keldysh~\cite{keldysh64,altland10,aoki14,sieberer16} formalisms.

In this Letter, we explore how this interplay affects long-range order. Focusing on the example of driven-dissipative superconductors, we  address  whether periodic driving enhances or reduces superconducting order. 
For static superconductors, the fitness criterion based on the commutator of the normal state Hamiltonian $\hat{H}_{0,\mathbf{k}}$ and the gap matrix $\hat{\Delta}_{\mathbf{k}}$  quantifies the potential stability  of superconducting states~\cite{ramires16,ramires18,footnote1}. It motivates us to propose a similar measure for  the driven system:
\begin{eqnarray}\label{eq:commutativity}
	\left[\hat{H}_{\pm1,\mathbf{k}},\hat{\Delta}_{\mathbf{k}}\right]_{-\zeta} = 0,
\end{eqnarray}
where $\hat{H}_{\pm1,\mathbf{k}}$ are the lowest order Fourier components of the drive Hamiltonian and ${\zeta=+1}$ (${\zeta=-1}$) denotes the commutator (anti-commutator).
Surprisingly, anti-commuting  drives re-order the  particle-hole structure of the Floquet spectral functions, as depicted in Fig.~\ref{fig:schematics}(a). When coupled to a thermal bath, this structure suggests a general scheme for the enhancement of the superconducting transition temperature, see Fig.~\ref{fig:schematics}(c).  Commuting drives, on the other hand,  are  generically detrimental to superconductivity. Our proposal is rather general and goes beyond standard mechanisms like dynamic squeezing of phonons~\cite{komnik16,knap16} and coherent destruction of electronic tunneling~\cite{sentef16,coulthard17,ido17}.
It opens the door for further studies for driven-dissipative engineering of exotic phases of matter in solid-state systems.

We consider a time-periodic single-particle Hamiltonian coupled to a  static bath
\begin{align}\label{eq:hamil}
	\mathcal{H}=& \sum_{\ell\in\mathbb{Z}}e^{i\ell\Omega t}\sum_{\mathbf{k}}\Psi_{\mathbf{k}}^\dagger H_{\ell,\mathbf{k}} \Psi_{\mathbf{k}} + \sum_{\mathbf{q},r} \xi_{\mathbf{q},r}b_{\mathbf{q},r}^\dagger b_{\mathbf{q},r}
	 \nonumber\\
	& + \sum_{\mathbf{k},\mathbf{q},s,r} \left(W_{\mathbf{k},\mathbf{q}} c^\dagger_{\mathbf{k},s} b_{\mathbf{q},r} + \mathrm{H.c.} \right).
\end{align}
where $c_{\mathbf{k},s}$ ($b_{\mathbf{q},r}$) are the fermionic annihilation operators of the system (bath) with momenta $\mathbf{k}$ ($\mathbf{q}$) and internal degrees of freedom $s$ ($r$), and $\Psi_{\mathbf{k}} = (c_{\mathbf{k},s_1},c_{\mathbf{k},s_2},\dots)^{\mathrm{T}}$;
$W_{\mathbf{k},\mathbf{q}}$ describes the weak system-bath coupling, while $\xi_{\mathbf{q},r}$ is the bath dispersion.  We use  the Floquet-Keldysh formalism in this Letter.
 
The Shirley-Floquet formalism captures the time periodicity of the Hamiltonian by expressing it in the Floquet representation as~\cite{shirley65,tsuji08,ivanov21} ($\hbar=k_B=1$)
\begin{eqnarray}\label{eq:time_periodic}
\mathbb{H}= \sum_{\ell\in\mathbb{Z}} H_{\ell} \mathbb{F}_{\ell} - \Omega \mathbb{N},
\end{eqnarray}
where $(\mathbb{F}_\ell)_{m,n} = \delta_{m,n-\ell}$ and $\mathbb{N}_{m,n} = m \delta_{m,n}$ are matrices in Floquet space, see Supplemental Material (SM)~\cite{supmat}.
Since Floquet matrix entries $(m,n)$ sharing the same $\ell=m-n$ are physically equivalent, they can be categorized into the $\ell$-th family of the Wigner representation~\cite{tsuji08} as
\begin{eqnarray}\label{eq:wigner}
	H_{\ell}(\omega) \equiv \mathbb{H}_{m,m+\ell}(\omega - (m+\ell/2)\Omega).
\end{eqnarray}
This structure also applies to other quantities as Green's functions $\mathbb{G}$ and  thermal distributions $\bbrho$.

In the Keldysh formalism, the bath is the source of two effects~\cite{altland10,aoki14,sieberer16}. First, the bath's spectral function $\Sigma(\omega)\equiv{\pi\sum_{\mathbf{q},r} |W_{\mathbf{k},\mathbf{q}}|^2 \delta(\omega-\xi_{\mathbf{q},r})}$ enters the system's retarded/advanced Green's function as $\omega\mapsto\omega_{\pm}=\omega\pm i\Sigma(\omega)$, widening its poles and rendering the excitation lifetime finite. Our results are qualitatively independent of the specific structure of $\Sigma(\omega)$, and we consider a Markovian bath $\Sigma(\omega)=\Sigma$ for an emphasis on the qualitative features.
Second, the bath defines the thermal density distribution $\bbrho$, which relates the retarded/advanced Green's function $\mathbb{G}^{R/A}=(\omega_{\pm} \mathbb{F}_0-\mathbb{H})^{-1}$, to the Keldysh Green's function through the generalized fluctuation-dissipation theorem in Floquet space~\cite{sieberer16}
\begin{eqnarray}\label{eq:green_function_keldysh}
	\mathbb{G}^{K} = \mathbb{G}^{R} \bbrho - \bbrho \mathbb{G}^{A}.
\end{eqnarray}
Particularly, in the lab frame where the bath is static, $\bbrho$ has a simple form in the Wigner representation as 
\begin{eqnarray}
	\rho_{\ell} = \tanh(\omega/2T) \delta_{\ell,0}.  
\end{eqnarray}

The drive induces an intrinsic relative rotation between the system and the bath, which cannot be eliminated in any rotating frame.
A usual strategy to solve the driven system is to diagonalize the system Hamiltonian in Floquet space, $\mathbb{H}\mapsto \mathbb{H}' = \mathbb{P}^\dagger \mathbb{H} \mathbb{P} = H_0' \mathbb{F}_0 - \Omega \mathbb{N}$, where $H_0'$ is the effective stroboscopic Hamiltonian, and the transformation operator $\mathbb{P}$ is a function of $\mathbb{F}_{\ell}$~\cite{ivanov21}. However, the frequency dependence in the Floquet structure of $\bbrho(\omega)$ makes it non-commuting with both $\mathbb{G}^{R}$ and $\mathbb{P}$. As a consequence, the bath distribution becomes generally non-thermal in the rotating frame, see SM~\cite{supmat},
\begin{eqnarray}\label{eq:intrinsic_bath_rotation}
	\bbrho \mapsto \bbrho' = \mathbb{P}^\dagger \bbrho \mathbb{P} \neq \bbrho.
\end{eqnarray}
For example, for systems effectively described by a Lindbladian, we obtain a frequency-detuned thermal distribution $\bbrho'(\omega) = \bbrho(\omega+\Omega)$, which suggests the inadequacy of the effective stroboscopic Hamiltonian to comprehensively capture the full dissipative physics. More generally, a Floquet Fermi liquid is obtained~\cite{shi23}, leading to more exotic effects.

From now on, we focus on the specific example of superconductors in the Nambu spinor basis ${\Psi_{\mathbf{k}} = (c_{\mathbf{k}\uparrow}, c_{\mathbf{k}\downarrow}, c_{-\mathbf{k}\uparrow}^\dagger, c_{-\mathbf{k}\downarrow}^\dagger)^\mathrm{T}}$. The single-particle Hamiltonian of the superconductor $\hat{H}_{\ell,\mathbf{k}}$ entering Eq.~\eqref{eq:hamil} is now explicitly written in Nambu (with Pauli matrices $\tau_i$) and spin ($\sigma_i$) spaces.
A time-independent attractive interaction ($g>0$) described by 
\begin{eqnarray}\label{eq:interaction}
	\mathcal{H}_\mathrm{int} = - \sum_{\mathbf{k}_1,\mathbf{k}_2}g_{\mathbf{k}_1,\mathbf{k}_2} c^\dagger_{\mathbf{k}_1\uparrow}c^\dagger_{-\mathbf{k}_1\downarrow}c_{\mathbf{k}_2\uparrow}c_{-\mathbf{k}_2\downarrow}
\end{eqnarray}
induces superconductivity in a specific symmetry channel characterized in the mean-field treatment by the spin-singlet gap matrix $\hat{\Delta}_{\mathbf{k}}=d_{\mathbf{k}} \tau_y\sigma_y$.

To evaluate the impact of drive and dissipation on the superconducting order, we treat the interaction in the mean-field limit using a Hubbard-Stratonovich transformation generalized to the Floquet-Keldysh space~\cite{yang21}. In contrast to Ref.~\cite{yang21}, we explicitly consider the effects stemming from Eqs.~\eqref{eq:green_function_keldysh} and \eqref{eq:intrinsic_bath_rotation}.
Representing the gap in terms of its Fourier components $\Delta(t) = \sum_{\ell\in\mathbb{Z}} \Delta_{\ell} e^{i\ell\Omega t}$, we obtain a self-consistent gap equation
\begin{eqnarray}\label{eq:gap_equation}
	\sum_{\ell\in\mathbb{Z}} \Delta_{\ell} \mathbb{F}_{\ell} = ig\int d\omega \sum_{\mathbf{k}} \tr_{\tau\sigma} \left(\hat{\Delta}_{\mathbf{k}} \hat{\mathbb{G}}^{K,}_{\mathbf{k}} \right),
\end{eqnarray}
where the trace is normalized $\tr \mathbf{1}= 1$, and the Keldysh Green's function is defined by Eq.~\eqref{eq:green_function_keldysh}, with
\begin{eqnarray}\label{eq:green_origin}
	\left(\hat{\mathbb{G}}^{R/A}_{\mathbf{k}}\right)^{-1}
	= \omega_{\pm} \tau_0\sigma_0\mathbb{F}_0 - \hat{\mathbb{H}}_{\mathbf{k}} - \hat{\Delta}_{\mathbf{k}}\sum_{\ell\in\mathbb{Z}} \Delta_{\ell} \mathbb{F}_{\ell}.
\end{eqnarray}
Notably, using ${[\mathbb{P},\mathbb{F}_\ell]=0}$ and the invariance of trace under similarity transformations, the gap equation can be shown to be invariant under a general change of reference frame, see SM~\cite{supmat}. 
This indicates that any gap oscillation in a driven-dissipative superconductor is intrinsic. It is a manifestation of Eq.~\eqref{eq:intrinsic_bath_rotation} and allows us to work in the lab frame where calculations are significantly simplified.

To investigate the effects of the intrinsic system-bath rotation, we consider a weak, sinusoidal drive with momentum-dependent amplitude $a_{\mathbf{k}}$ implemented upon a superconductor with electronic dispersion $\epsilon_\mathbf{k}$,
\begin{eqnarray}\label{eq:drive}
	\hat{H}_{0,\mathbf{k}}= \epsilon_\mathbf{k} \tau_z\sigma_0,\quad \hat{H}_{\pm1,\mathbf{k}} = \frac{a_\mathbf{k}}{2} \tau_\mu \sigma_\nu, \quad \hat{H}_{|\ell|\ge2}=0,
\end{eqnarray}
with $\mu\in \{0,z\}$ and $|a_{\mathbf{k}}|\ll \Omega$ . The driving frequency is at least comparable to the electronic energy scale $\Omega\gtrsim \epsilon_{\mathbf{k}}$.
The weak drive motivates us to assume a priori that the gap is dominated by its time-averaged value $\Delta_0\gg|\Delta_{|\ell|\ge1}|$,
\begin{eqnarray}\label{eq:assumption}
	\hat{\Delta}_{\mathbf{k}}\sum_{\ell\in\mathbb{Z}} \Delta_{\ell} \mathbb{F}_{\ell} \approx \Delta_0 d_{\mathbf{k}} \tau_y\sigma_y\mathbb{F}_0,
\end{eqnarray}
where we have chosen the gauge $\Delta_0\ge0$.
This approximation captures the dominant features of the driven-dissipative superconductor by allowing us to write a Dyson equation for the Green's function as
\begin{eqnarray}\label{eq:G_expansion}
	\hat{\mathbb{G}}^{R}_{\mathbf{k}} = \hat{\widetilde{\mathbb{G}}}{\mathstrut}^R_{\mathbf{k}} \sum_{n=0}^{\infty}\left[ \frac{a_{\mathbf{k}}}{2}\tau_\mu \sigma_\nu(\mathbb{F}_1+\mathbb{F}_{-1}) \hat{\widetilde{\mathbb{G}}}{\mathstrut}^R_{\mathbf{k}}\right]^n.
\end{eqnarray}
Here, $\hat{\widetilde{\mathbb{G}}}{\mathstrut}^R_{\mathbf{k}}\equiv \hat{\mathbb{G}}_{\mathbf{k}}^{R}(a_{\mathbf{k}}=0)$ describes the non-driven system and can be solved as $\left(\hat{\widetilde{G}}{\mathstrut}^R_{\mathbf{k}}\right)_{\ell}(\omega) = \delta_{\ell,0}\hat{\widetilde{G}}{\mathstrut}^R_{\mathbf{k}}(\omega)$, with
\begin{eqnarray}\label{eq:G_static}
	\hat{\widetilde{G}}{\mathstrut}^R_{\mathbf{k}}(\omega) = \frac{\omega_+ \tau_0\sigma_0 + \epsilon_{\mathbf{k}} \tau_z\sigma_0 + \Delta_0 d_{\mathbf{k}} \tau_y\sigma_y}{\omega_+^2-E_{\mathbf{k}}^2},
\end{eqnarray}
where $E_\mathbf{k}=\sqrt{\epsilon_\mathbf{k}^2+\Delta_0^2 d_{\mathbf{k}}^2}$.
By substituting the ansatz Eq.~\eqref{eq:G_expansion} into the gap equation Eq.~\eqref{eq:gap_equation}, the Fourier components of the gap $\Delta_{\ell}$ are determined by the Wigner Green's functions ${\left(\hat{G}_{\mathbf{k}}^{R}\right)_{\ell}(\omega) \sim \left(\frac{a_{\mathbf{k}}}{2}\right)^{|\ell|}\hat{\widetilde{G}}{\mathstrut}^R_{\mathbf{k}}(\omega) \prod_{n=1}^{\ell}\left[\tau_\mu\sigma_\nu\hat{\widetilde{G}}{\mathstrut}^R_{\mathbf{k}}(\omega\pm n \Omega)\right]}$.
By noticing that $a_{\mathbf{k}}\hat{\widetilde{G}}{\mathstrut}^R_{\mathbf{k}}(E_\mathbf{k}-n\Omega)\sim a_{\mathbf{k}}/\Omega$ for $n\neq0$, we find $\Delta_{\ell} \sim \left(\frac{a_{\mathbf{k}}}{\Omega}\right)^{|\ell|}\Delta_0$, which validates Eq.~\eqref{eq:assumption}. 

The dominance of the static component of the order parameter indicates that the superconducting excitations can be characterized by the electron-hole spectral function $\mathcal{A}^{\mathrm{eh}}$, the anomalous spectral function $\mathcal{A}^{\mathrm{an}}$, and the anomalous response function $\mathcal{R}^{\mathrm{an}}$ of the $\ell=0$ Wigner Green's functions,
\begin{align}\label{eq:def_functions}
	\begin{split}
		\mathcal{A}^{\mathrm{eh}}_{\mathbf{k}}(\omega) &= -\Im\tr_{\tau\sigma} \left[\left(\hat{G}_{\mathbf{k}}^{R}\right)_{\ell= 0}\right]/\pi \\
		\mathcal{A}^{\mathrm{an}}_{\mathbf{k}}(\omega) &= -\Im\tr_{\tau\sigma} \left[d_\mathbf{k}\tau_y\sigma_y\left(\hat{G}_{\mathbf{k}}^{R}\right)_{\ell= 0}\right]/\pi \\
		\mathcal{R}^{\mathrm{an}}_{\mathbf{k}}(\omega) &= -\Im\tr_{\tau\sigma} \left[d_\mathbf{k}\tau_y\sigma_y\left(\hat{G}_{\mathbf{k}}^{K}\right)_{\ell= 0}\right]/\pi.
	\end{split}
\end{align}
Notably, the spectral functions capture the driving effects on the superconductor, while the response function captures the thermal effects induced by the bath. The gap equation Eq.~\eqref{eq:gap_equation} is now reduced to
\begin{eqnarray}\label{eq:gap_equation_Delta_0}
	\Delta_0 = ig\sum_{\mathbf{k}} \int d\omega \mathcal{R}^{\mathrm{an}}_\mathbf{k}(\omega),
\end{eqnarray}
where Eq.~\eqref{eq:green_function_keldysh} manifests in the lab frame as,
\begin{eqnarray}\label{eq:relation_R_and_A}
	\mathcal{R}^{\mathrm{an}}_{\mathbf{k}}(\omega) = \tanh\left(\frac{\omega}{2T}\right)\mathcal{A}^{\mathrm{an}}_{\mathbf{k}}(\omega).
\end{eqnarray}

\begin{figure}[t]
	\centering
	\includegraphics[width=\columnwidth]{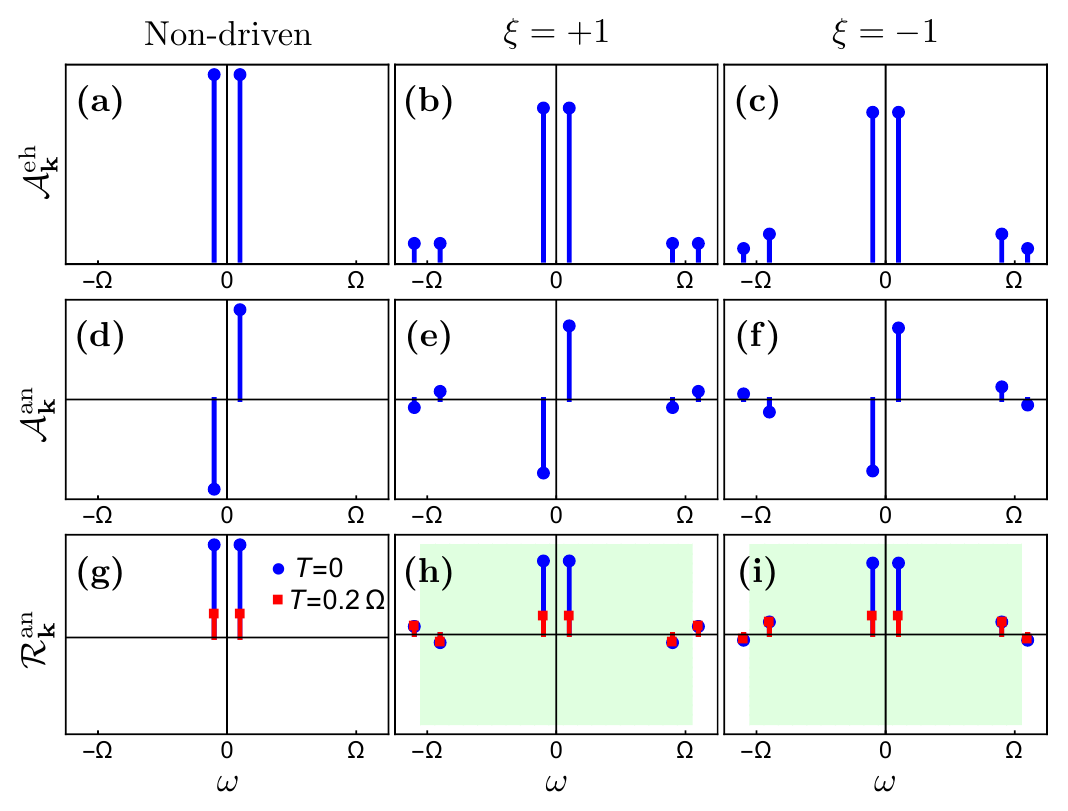}
	\caption{
		(a-c) The electron-hole spectral functions, (d-f) the anomalous spectral functions, and (g-i) the anomalous response functions [Eq.~\eqref{eq:def_functions}] of the weakly-dissipative $\Sigma\to 0^+$ superconductors in the first three temporal Brillouin zones $\omega\in(-3\Omega/2,3\Omega/2]$, for the (a,d,g) undriven, (b,e,h) commuting ($\zeta=+1$), and (c,f,i) anti-commuting ($\zeta=-1$) cases. Parameters are taken as $\epsilon_{\mathbf{k}}=0$, $\Delta_0 d_{\mathbf{k}}= 0.1\Omega$, $a_{\mathbf{k}}=0.6\Omega$. (g-i) For the response functions, results for zero temperature (blue dots) and intermediate temperature (red squares) are shown. 
		The green background indicates the region below the cutoff frequency $\Omega_C$.
	}
	\label{fig:spectral_response}
\end{figure}

The driving induces a specific Floquet structure in the spectral functions. For the non-driven system, $\left(\hat{\widetilde{G}}{\mathstrut}^R_{\mathbf{k}}\right)_{\ell=0}$ has two equivalent poles at ${\omega_+=\pm E_{\mathbf{k}}}$ reflecting the particle-hole symmetry imposed by the Nambu basis. The corresponding spectral functions are [Fig.~\ref{fig:spectral_response}(a,d)]
\begin{align}
	\begin{split}
		\widetilde{\mathcal{A}}^{\mathrm{eh}}_{\mathbf{k}}(\omega) &= \frac{1}{2} \left[L_{\Sigma}(\omega-E_{\mathbf{k}}) + L_{\Sigma}(\omega+E_{\mathbf{k}})\right] \\
		\widetilde{\mathcal{A}}^{\mathrm{an}}_{\mathbf{k}}(\omega) &= \frac{\Delta_0 d_{\mathbf{k}}}{2E_{\mathbf{k}}}\left[L_{\Sigma}(\omega-E_{\mathbf{k}}) - L_{\Sigma}(\omega+E_{\mathbf{k}})\right],
	\end{split}
\end{align}
where $L_{\Sigma}$ denotes a Lorentzian distribution $L_\Sigma(\omega)=\frac{\Sigma}{\pi(\omega^2+\Sigma^2)}$, which recovers a Dirac delta distribution $L_{\Sigma\to 0^+}(\omega) = \delta(\omega)$ in the weakly dissipative limit.
In contrast, the Green's function for the driven system $\left(\hat{G}_{\mathbf{k}}^{R}\right)_{\ell=0}$ acquires infinitely many shifted poles at $\omega_{+}=\pm E_\mathbf{k}+n\Omega$, $n\in\mathbb{Z}$. In the high-frequency limit $\Omega\gg\epsilon_{\mathbf{k}}$ for clarity, we find the lowest-order contributions to the spectral functions in the $(n+1)$-th temporal Brillouin zone $\omega/\Omega\in(n-1/2,n+1/2]$ behave as [Fig.~\ref{fig:spectral_response}(b,c,e,f)]
\begin{align}\label{eq:spectral_functions_approx}
	\begin{split}
	\mathcal{A}^{\mathrm{eh}}_{\mathbf{k}}(\omega) \approx \sum_{n=-\infty}^{\infty}\frac{1}{(n!)^2}\left(\frac{a_{\mathbf{k}}}{2\Omega}\right)^{2|n|} \widetilde{\mathcal{A}}^{\mathrm{eh}}_{\mathbf{k}}(\omega-n\Omega) \\ 
	\mathcal{A}^{\mathrm{an}}_{\mathbf{k}}(\omega) \approx \sum_{n=-\infty}^{\infty}\frac{\zeta^n}{(n!)^2}\left(\frac{a_{\mathbf{k}}}{2\Omega}\right)^{2|n|} \widetilde{\mathcal{A}}^{\mathrm{an}}_{\mathbf{k}}(\omega-n\Omega). 
	\end{split}
\end{align}
The following discussion is valid for large $\Omega$ up to $\Omega\approx2 E_{\mathbf{k}}$, in which case the drive induces a parametric resonance between the particle and hole excitations~\cite{supmat}.

We now elucidate the role of the commutator, Eq.~\eqref{eq:commutativity}.
In the commuting case [cf.\ Eq.~\eqref{eq:commutativity}], ${\zeta=+1}$, e.g., when the drive is realized by a magnetic field along the $y$ direction $\mu=0$, $\nu=y$, the Wigner Green's functions admit analytical solutions as
$\left(\hat{G}_{\mathbf{k}}^{R}\right)_{\ell}(\omega)
= \sum_{\tilde{\ell}\in \mathbb{Z}+\ell/2} \hat{\widetilde{G}}{\mathstrut}^R_{\mathbf{k}}\left(\omega+\tilde{\ell}\Omega\right)J_{\tilde{\ell}+\ell/2}\left(\frac{a_{\mathbf{k}}}{\Omega}\right)J_{\tilde{\ell}-\ell/2}\left(\frac{a_{\mathbf{k}}}{\Omega}\right)$,
with $J_{\ell}$ the Bessel functions of the first kind, see SM~\cite{supmat}. 
The physical consequences of the driving are not reflected in the spectral functions $\int_{-\infty}^{\infty} d\omega \mathcal{A}^{\mathrm{eh}/\mathrm{an}}_{\mathbf{k}}(\omega) = \int_{-\infty}^{\infty} d\omega \widetilde{\mathcal{A}}^{\mathrm{eh}/\mathrm{an}}_{\mathbf{k}}(\omega)$~\cite{uhrig19}, but only in the response function $\int_{-\infty}^{\infty} d\omega \mathcal{R}^{\mathrm{an}}_{\mathbf{k}}(\omega) < \int_{-\infty}^{\infty} d\omega \widetilde{\mathcal{R}}^{\mathrm{an}}_{\mathbf{k}}(\omega)$.  In this case, the intrinsic system-bath rotation is detrimental to superconductivity, cf.\ Eq.~\eqref{eq:gap_equation_Delta_0}, as confirmed in Fig.~\eqref{fig:schematics}(b).

In contrast, the anti-commuting case, $\zeta=-1$, has more intriguing features. This can be realized by ${\mu=z,\,\nu=y}$, associated to a Rashba-like spin-orbit coupling. In the anomalous spectral function $\mathcal{A}^{\mathrm{an}}$ [Fig.~\ref{fig:spectral_response}(f)], an extra phase shift of $\pi$ is introduced in the gap $\Delta_0\to-\Delta_0$ as one moves from one temporal Brillouin zone to the next, which results in an alternating sign in $\mathcal{A}^{\mathrm{an}}$ across temporal Brillouin zones. Effectively, the roles of particle and hole are successively exchanged, as depicted in Fig.~\ref{fig:schematics}(a).
The physical consequences of this phase shifting is evident in the response function $\mathcal{R}_{\mathrm{an}}$ [Fig.~\ref{fig:spectral_response}(i)]. Its four poles closest to the Fermi surface at $\omega_+=\pm E_{\mathbf{k}}$ and $\omega_+=\pm (E_{\mathbf{k}}-\Omega)$ carry positive weights. 

We now discuss the effects of an anti-commuting drive on the gap equation,  Eq.~\eqref{eq:gap_equation_Delta_0}. To the lowest order, the response functions at $\omega_+=\pm E_{\mathbf{k}}$ contribute to the gap with $\frac{\Delta_0}{E_{\mathrm{k}}}\tanh\left(\frac{E_{\mathrm{k}}}{2T}\right)$, cf.\ Eq.~\eqref{eq:G_expansion}. This contribution is significant at zero temperature and very sensitive to temperature because of its proximity to the Fermi surface. In contrast, the response functions at $\omega_+=\pm (E_{\mathbf{k}}-\Omega)$ contribute with $\frac{a^2}{4\Omega^2}\frac{\Delta_0(\Omega^2-4\epsilon_{\mathbf{k}}^2)}{E_{\mathrm{k}}(\Omega-2E_{\mathrm{k}})^2}\tanh\left(\frac{\Omega-E_{\mathrm{k}}}{2T}\right)$. Particularly, at large driving frequency $\Omega\gg E_{\mathbf{k}}$, these excitations are distant from the Fermi surface, and their contribution is much less sensitive to temperature. Consequently, the combined contribution from all four excitations is expected to be greater than the non-driven counterpart [Fig.~\ref{fig:spectral_response}(g)] at high temperatures $T>E_{\mathbf{k}}$.
To enhance superconductivity, the analysis of the response function motivates us to suppress all other excitations, particularly the ones detrimental to the gap at $\omega_+=\pm (E_{\mathbf{k}}+\Omega)$. This can be achieved by choosing  a drive close resonance to the intrinsic interaction's cutoff frequency $\Omega_C$, such that ${\delta\Omega \equiv \Omega-\Omega_C\ll\Omega}$. We assume that the cutoff, as given, for example, by the phonon Debye frequency, remains unaffected by the drive.

We show how our scheme works for  the simple scenario of a flat-band system with $s$-wave superconductivity and a drive associated with a Rashba-like spin-orbit coupling, i.e.,
\begin{eqnarray}\label{eq:flat_band_hamil}
	\epsilon_\mathbf{k}=0, \quad d_{\mathbf{k}}=1, \quad a_{\mathbf{k}}=a \sgn(k_x).
\end{eqnarray}
To the order of $O(a^2/\Omega^2)$, the anomalous spectral function can be approximated by
\begin{eqnarray}\label{eq:response_function}
		\mathcal{A}^{\mathrm{an}}(\omega)
		&\approx&
		\left(\frac{1}{2}-\frac{a^2}{4\Omega^2}\right)[L_\Sigma(\omega-\Delta_0)-L_\Sigma(\omega+\Delta_0)] \\
		&& +\frac{a^2}{8\Omega^2}[L_\Sigma(\omega-\Omega+\Delta_0) - L_\Sigma(\omega+\Omega-\Delta_0)], \nonumber
\end{eqnarray}
where excitations lying beyond the cutoff frequency $|\omega|>\Omega_C$ are omitted. 
The corresponding gap equation for $\Delta_0$ in the weakly dissipative limit $\Sigma\to 0^+$ now reads
\begin{align}\label{eq:sc_eqn_analytic}
	\begin{split}
		\frac{\Delta_0}{gN} \approx & \left(1-\frac{a^2}{2\Omega^2}\right)\tanh\left(\frac{\Delta_0}{2T}\right) \\ 
		& +\frac{a^2}{4\Omega^2}\tanh\left(\frac{\Omega}{2T}\right), \quad \forall \Delta_0>\delta\Omega,
	\end{split}
\end{align}
where $N$ is the total density of states. Dissipation with characteristic energies comparable to the gap $\Delta_0$ will suppress superconductivity~\cite{yang21}.

The enhancement of superconductivity can be confirmed by a finite order parameter beyond the critical temperature of the non-driven superconductor $T_{c0}=gN/2$. At high temperatures ${T \gg \Omega}$, the gap of the driven superconductor exhibits an asymptotic behavior of $\Delta_0\sim 1/T$,
with a first-order jump from finite to vanishing values at the temperature
\begin{eqnarray}\label{eq:flat_temp_increase}
	\frac{T_c}{T_{c0}} \approx  \frac{a^2}{8\Omega^2}\frac{\Omega}{\delta\Omega}.
\end{eqnarray}
This result implies the existence of superconductivity at arbitrarily high temperatures for a resonantly tuned driving frequency $\delta\Omega\to0$. 
We confirm our analysis by numerically solving Eq.~\eqref{eq:gap_equation} for flat-band superconductors Eq.~\eqref{eq:flat_band_hamil} at different temperatures; see Fig.~\ref{fig:schematics}(c).
The numerical solutions confirm our assumption that the oscillating components are negligible compared to the constant component, see SM~\cite{supmat}.
In  the weak coupling limit, the robustness of our scheme manifests itself in its qualitative validity i) when the driving frequency is set to be resonant with any odd submultiple of the cutoff frequency, and ii) for dispersive superconductors (see SM~\cite{supmat}) and superconducting gap of symmetries other than $s$-wave.

In summary, the Floquet-Keldysh formalism, in conjunction with a generalization of the fitness criterion to driven superconductors, provides a framework to engineer specific drives to enhance the superconducting transition temperature~\cite{footnote1}.
In the static limit, a normal state fully anti-commuting with the gap matrix ${[\hat{H}_{\ell=0,\mathbf{k}},\hat{\Delta}_{\mathbf{k}}]_{+}=0}$ induces a maximal gap, which can be reduced by any commutativity. Based on a given static system, a drive in the normal state anti-commuting with the gap matrix ${[\hat{H}_{\ell=\pm1,\mathbf{k}},\hat{\Delta}_{\mathbf{k}}]_{+}=0}$ further potentially enhances the gap in the vicinity of the transition temperature of the static system.
Our mean-field treatment of the interaction mediator intrinsically neglects its fluctuations. These fluctuations potentially stabilize the superconductivity in equilibrium~\cite{chubukov20,marsiglio20,you19,liu22,chakraborty21,kelly22}, but can also induce thermalization and quasiparticle scattering in driven systems. These effects should be further investigated in the future.

We clarify the distinctions between our scheme with others in literature.
Schemes for the enhancement of superconducting order based on phonon driving and squeezing~\cite{komnik16,knap16} or coherent suppression of electron tunneling~\cite{sentef16,coulthard17,ido17}  rely on the description of stroboscopic Hamiltonians. Moreover, Ref.~\cite{hart19} studied a Rabi drive using the formalism presented here.  Unlike the sinusoidal drive discussed here, the Rabi drive inherently induces solely co-rotating dynamics, validating a simplified treatment using Lindbladians, cf.\ Eq.~\eqref{eq:intrinsic_bath_rotation}.
Furthermore, the interaction mediator also acts as a thermal bath, and it is usually coupled to the electronic system in a fundamentally different form than Eq.~\eqref{eq:hamil}. Its effects have been investigated for cavity-mediated~\cite{chakraborty21} and driven phonon-mediated~\cite{eliashberg71,ivlev73} superconductors. Particularly in the latter case, features similar to our systems are observed, like first-order transition and the associated hysteresis of the superconducting gap.

Our results indicate that the enhancement of the superconducting transition temperature is most pronounced for flat-band systems~\cite{supmat}. Material platforms with flat electronic bands have been extensively discussed in the context of twisted two-dimensional van der Waals materials~\cite{kennes21} and heavy fermion systems~\cite{ramires22}. Flat bands have also been theoretically identified near the Fermi level in three-dimensional materials, including Weyl-Kondo semimetals~\cite{lai18} and materials with Kagome, pyrochlore, or Lieb sublattice structures~\cite{regnault22}. In addition, recent experimental developments have shown that it is possible to generate Floquet bands in van der Waals materials~\cite{wang13}  and graphene~\cite{mciver20} with light in the THz regime. These results suggest that the necessary experimental ingredients for realizing our proposal are readily available. Moreover, materials with other unconventional band structures like van Hove singularities, which has high density of states close to the Fermi surface, can also potentially be candidates of our proposal and manifest more intriguing features~\cite{shi23}.

To conclude, inspired by the interplay between drive and dissipation in quantum optical systems, we have illustrated the capability of driven-dissipative engineering, specifically in tailoring superconducting order. This stimulates exploration and generalization of the technique to other ordered states of matter in solid state platforms.

\begin{acknowledgments}
	R.L. and R.C.  acknowledge funding from the ETH Grant, and A.R. acknowledges financial support from the Swiss National Science Foundation (SNSF) through an Ambizione Grant No.\ 186043.
\end{acknowledgments}

\bibliographystyle{apsrev}
\bibliography{References}

\begin{thebibliography}{68}
\expandafter\ifx\csname natexlab\endcsname\relax\def\natexlab#1{#1}\fi
\expandafter\ifx\csname bibnamefont\endcsname\relax
  \def\bibnamefont#1{#1}\fi
\expandafter\ifx\csname bibfnamefont\endcsname\relax
  \def\bibfnamefont#1{#1}\fi
\expandafter\ifx\csname citenamefont\endcsname\relax
  \def\citenamefont#1{#1}\fi
\expandafter\ifx\csname url\endcsname\relax
  \def\url#1{\texttt{#1}}\fi
\expandafter\ifx\csname urlprefix\endcsname\relax\def\urlprefix{URL }\fi
\providecommand{\bibinfo}[2]{#2}
\providecommand{\eprint}[2][]{\url{#2}}

\bibitem[{\citenamefont{Komnik and Thorwart}(2016)}]{komnik16}
\bibinfo{author}{\bibfnamefont{A.}~\bibnamefont{Komnik}} \bibnamefont{and}
  \bibinfo{author}{\bibfnamefont{M.}~\bibnamefont{Thorwart}},
  \bibinfo{journal}{The European Physical Journal B}
  \textbf{\bibinfo{volume}{89}}, \bibinfo{pages}{244} (\bibinfo{year}{2016}),
  \urlprefix\url{https://doi.org/10.1140/epjb/e2016-70528-1}.

\bibitem[{\citenamefont{Knap et~al.}(2016)\citenamefont{Knap, Babadi, Refael,
  Martin, and Demler}}]{knap16}
\bibinfo{author}{\bibfnamefont{M.}~\bibnamefont{Knap}},
  \bibinfo{author}{\bibfnamefont{M.}~\bibnamefont{Babadi}},
  \bibinfo{author}{\bibfnamefont{G.}~\bibnamefont{Refael}},
  \bibinfo{author}{\bibfnamefont{I.}~\bibnamefont{Martin}}, \bibnamefont{and}
  \bibinfo{author}{\bibfnamefont{E.}~\bibnamefont{Demler}},
  \bibinfo{journal}{Phys. Rev. B} \textbf{\bibinfo{volume}{94}},
  \bibinfo{pages}{214504} (\bibinfo{year}{2016}),
  \urlprefix\url{https://link.aps.org/doi/10.1103/PhysRevB.94.214504}.

\bibitem[{\citenamefont{Sentef et~al.}(2016)\citenamefont{Sentef, Kemper,
  Georges, and Kollath}}]{sentef16}
\bibinfo{author}{\bibfnamefont{M.~A.} \bibnamefont{Sentef}},
  \bibinfo{author}{\bibfnamefont{A.~F.} \bibnamefont{Kemper}},
  \bibinfo{author}{\bibfnamefont{A.}~\bibnamefont{Georges}}, \bibnamefont{and}
  \bibinfo{author}{\bibfnamefont{C.}~\bibnamefont{Kollath}},
  \bibinfo{journal}{Phys. Rev. B} \textbf{\bibinfo{volume}{93}},
  \bibinfo{pages}{144506} (\bibinfo{year}{2016}),
  \urlprefix\url{https://link.aps.org/doi/10.1103/PhysRevB.93.144506}.

\bibitem[{\citenamefont{Coulthard et~al.}(2017)\citenamefont{Coulthard, Clark,
  Al-Assam, Cavalleri, and Jaksch}}]{coulthard17}
\bibinfo{author}{\bibfnamefont{J.~R.} \bibnamefont{Coulthard}},
  \bibinfo{author}{\bibfnamefont{S.~R.} \bibnamefont{Clark}},
  \bibinfo{author}{\bibfnamefont{S.}~\bibnamefont{Al-Assam}},
  \bibinfo{author}{\bibfnamefont{A.}~\bibnamefont{Cavalleri}},
  \bibnamefont{and} \bibinfo{author}{\bibfnamefont{D.}~\bibnamefont{Jaksch}},
  \bibinfo{journal}{Phys. Rev. B} \textbf{\bibinfo{volume}{96}},
  \bibinfo{pages}{085104} (\bibinfo{year}{2017}),
  \urlprefix\url{https://link.aps.org/doi/10.1103/PhysRevB.96.085104}.

\bibitem[{\citenamefont{Ido et~al.}(2017)\citenamefont{Ido, Ohgoe, and
  Imada}}]{ido17}
\bibinfo{author}{\bibfnamefont{K.}~\bibnamefont{Ido}},
  \bibinfo{author}{\bibfnamefont{T.}~\bibnamefont{Ohgoe}}, \bibnamefont{and}
  \bibinfo{author}{\bibfnamefont{M.}~\bibnamefont{Imada}},
  \bibinfo{journal}{Science Advances} \textbf{\bibinfo{volume}{3}},
  \bibinfo{pages}{e1700718} (\bibinfo{year}{2017}),
  \urlprefix\url{https://www.science.org/doi/abs/10.1126/sciadv.1700718}.

\bibitem[{\citenamefont{Mitrano et~al.}(2016)\citenamefont{Mitrano, Cantaluppi,
  Nicoletti, Kaiser, Perucchi, Lupi, Di~Pietro, Pontiroli, Ricc{\`o}, Clark
  et~al.}}]{mitrano16}
\bibinfo{author}{\bibfnamefont{M.}~\bibnamefont{Mitrano}},
  \bibinfo{author}{\bibfnamefont{A.}~\bibnamefont{Cantaluppi}},
  \bibinfo{author}{\bibfnamefont{D.}~\bibnamefont{Nicoletti}},
  \bibinfo{author}{\bibfnamefont{S.}~\bibnamefont{Kaiser}},
  \bibinfo{author}{\bibfnamefont{A.}~\bibnamefont{Perucchi}},
  \bibinfo{author}{\bibfnamefont{S.}~\bibnamefont{Lupi}},
  \bibinfo{author}{\bibfnamefont{P.}~\bibnamefont{Di~Pietro}},
  \bibinfo{author}{\bibfnamefont{D.}~\bibnamefont{Pontiroli}},
  \bibinfo{author}{\bibfnamefont{M.}~\bibnamefont{Ricc{\`o}}},
  \bibinfo{author}{\bibfnamefont{S.~R.} \bibnamefont{Clark}},
  \bibnamefont{et~al.}, \bibinfo{journal}{Nature}
  \textbf{\bibinfo{volume}{530}}, \bibinfo{pages}{461} (\bibinfo{year}{2016}),
  \urlprefix\url{https://doi.org/10.1038/nature16522}.

\bibitem[{\citenamefont{Kennes et~al.}(2019)\citenamefont{Kennes, Claassen,
  Sentef, and Karrasch}}]{kennes19}
\bibinfo{author}{\bibfnamefont{D.~M.} \bibnamefont{Kennes}},
  \bibinfo{author}{\bibfnamefont{M.}~\bibnamefont{Claassen}},
  \bibinfo{author}{\bibfnamefont{M.~A.} \bibnamefont{Sentef}},
  \bibnamefont{and} \bibinfo{author}{\bibfnamefont{C.}~\bibnamefont{Karrasch}},
  \bibinfo{journal}{Phys. Rev. B} \textbf{\bibinfo{volume}{100}},
  \bibinfo{pages}{075115} (\bibinfo{year}{2019}),
  \urlprefix\url{https://link.aps.org/doi/10.1103/PhysRevB.100.075115}.

\bibitem[{\citenamefont{Buzzi et~al.}(2020)\citenamefont{Buzzi, Nicoletti,
  Fechner, Tancogne-Dejean, Sentef, Georges, Biesner, Uykur, Dressel, Henderson
  et~al.}}]{buzzi20}
\bibinfo{author}{\bibfnamefont{M.}~\bibnamefont{Buzzi}},
  \bibinfo{author}{\bibfnamefont{D.}~\bibnamefont{Nicoletti}},
  \bibinfo{author}{\bibfnamefont{M.}~\bibnamefont{Fechner}},
  \bibinfo{author}{\bibfnamefont{N.}~\bibnamefont{Tancogne-Dejean}},
  \bibinfo{author}{\bibfnamefont{M.~A.} \bibnamefont{Sentef}},
  \bibinfo{author}{\bibfnamefont{A.}~\bibnamefont{Georges}},
  \bibinfo{author}{\bibfnamefont{T.}~\bibnamefont{Biesner}},
  \bibinfo{author}{\bibfnamefont{E.}~\bibnamefont{Uykur}},
  \bibinfo{author}{\bibfnamefont{M.}~\bibnamefont{Dressel}},
  \bibinfo{author}{\bibfnamefont{A.}~\bibnamefont{Henderson}},
  \bibnamefont{et~al.}, \bibinfo{journal}{Phys. Rev. X}
  \textbf{\bibinfo{volume}{10}}, \bibinfo{pages}{031028}
  (\bibinfo{year}{2020}),
  \urlprefix\url{https://link.aps.org/doi/10.1103/PhysRevX.10.031028}.

\bibitem[{\citenamefont{Budden et~al.}(2021)\citenamefont{Budden, Gebert,
  Buzzi, Jotzu, Wang, Matsuyama, Meier, Laplace, Pontiroli, Ricc{\`o}
  et~al.}}]{budden21}
\bibinfo{author}{\bibfnamefont{M.}~\bibnamefont{Budden}},
  \bibinfo{author}{\bibfnamefont{T.}~\bibnamefont{Gebert}},
  \bibinfo{author}{\bibfnamefont{M.}~\bibnamefont{Buzzi}},
  \bibinfo{author}{\bibfnamefont{G.}~\bibnamefont{Jotzu}},
  \bibinfo{author}{\bibfnamefont{E.}~\bibnamefont{Wang}},
  \bibinfo{author}{\bibfnamefont{T.}~\bibnamefont{Matsuyama}},
  \bibinfo{author}{\bibfnamefont{G.}~\bibnamefont{Meier}},
  \bibinfo{author}{\bibfnamefont{Y.}~\bibnamefont{Laplace}},
  \bibinfo{author}{\bibfnamefont{D.}~\bibnamefont{Pontiroli}},
  \bibinfo{author}{\bibfnamefont{M.}~\bibnamefont{Ricc{\`o}}},
  \bibnamefont{et~al.}, \bibinfo{journal}{Nature Physics}
  \textbf{\bibinfo{volume}{17}}, \bibinfo{pages}{611} (\bibinfo{year}{2021}),
  \urlprefix\url{https://doi.org/10.1038/s41567-020-01148-1}.

\bibitem[{\citenamefont{Takasan et~al.}(2017)\citenamefont{Takasan, Daido,
  Kawakami, and Yanase}}]{takasan17}
\bibinfo{author}{\bibfnamefont{K.}~\bibnamefont{Takasan}},
  \bibinfo{author}{\bibfnamefont{A.}~\bibnamefont{Daido}},
  \bibinfo{author}{\bibfnamefont{N.}~\bibnamefont{Kawakami}}, \bibnamefont{and}
  \bibinfo{author}{\bibfnamefont{Y.}~\bibnamefont{Yanase}},
  \bibinfo{journal}{Phys. Rev. B} \textbf{\bibinfo{volume}{95}},
  \bibinfo{pages}{134508} (\bibinfo{year}{2017}),
  \urlprefix\url{https://link.aps.org/doi/10.1103/PhysRevB.95.134508}.

\bibitem[{\citenamefont{Zhang and Das~Sarma}(2021)}]{zhang21}
\bibinfo{author}{\bibfnamefont{R.-X.} \bibnamefont{Zhang}} \bibnamefont{and}
  \bibinfo{author}{\bibfnamefont{S.}~\bibnamefont{Das~Sarma}},
  \bibinfo{journal}{Phys. Rev. Lett.} \textbf{\bibinfo{volume}{127}},
  \bibinfo{pages}{067001} (\bibinfo{year}{2021}),
  \urlprefix\url{https://link.aps.org/doi/10.1103/PhysRevLett.127.067001}.

\bibitem[{\citenamefont{Dehghani et~al.}(2021)\citenamefont{Dehghani, Hafezi,
  and Ghaemi}}]{dehghani21}
\bibinfo{author}{\bibfnamefont{H.}~\bibnamefont{Dehghani}},
  \bibinfo{author}{\bibfnamefont{M.}~\bibnamefont{Hafezi}}, \bibnamefont{and}
  \bibinfo{author}{\bibfnamefont{P.}~\bibnamefont{Ghaemi}},
  \bibinfo{journal}{Phys. Rev. Res.} \textbf{\bibinfo{volume}{3}},
  \bibinfo{pages}{023039} (\bibinfo{year}{2021}),
  \urlprefix\url{https://link.aps.org/doi/10.1103/PhysRevResearch.3.023039}.

\bibitem[{\citenamefont{Kitamura and Aoki}(2022)}]{kitamura22}
\bibinfo{author}{\bibfnamefont{S.}~\bibnamefont{Kitamura}} \bibnamefont{and}
  \bibinfo{author}{\bibfnamefont{H.}~\bibnamefont{Aoki}},
  \bibinfo{journal}{Communications Physics} \textbf{\bibinfo{volume}{5}},
  \bibinfo{pages}{174} (\bibinfo{year}{2022}),
  \urlprefix\url{https://doi.org/10.1038/s42005-022-00936-w}.

\bibitem[{\citenamefont{Triola and Balatsky}(2016)}]{triola16}
\bibinfo{author}{\bibfnamefont{C.}~\bibnamefont{Triola}} \bibnamefont{and}
  \bibinfo{author}{\bibfnamefont{A.~V.} \bibnamefont{Balatsky}},
  \bibinfo{journal}{Phys. Rev. B} \textbf{\bibinfo{volume}{94}},
  \bibinfo{pages}{094518} (\bibinfo{year}{2016}),
  \urlprefix\url{https://link.aps.org/doi/10.1103/PhysRevB.94.094518}.

\bibitem[{\citenamefont{Cayao et~al.}(2021)\citenamefont{Cayao, Triola, and
  Black-Schaffer}}]{cayao21}
\bibinfo{author}{\bibfnamefont{J.}~\bibnamefont{Cayao}},
  \bibinfo{author}{\bibfnamefont{C.}~\bibnamefont{Triola}}, \bibnamefont{and}
  \bibinfo{author}{\bibfnamefont{A.~M.} \bibnamefont{Black-Schaffer}},
  \bibinfo{journal}{Phys. Rev. B} \textbf{\bibinfo{volume}{103}},
  \bibinfo{pages}{104505} (\bibinfo{year}{2021}),
  \urlprefix\url{https://link.aps.org/doi/10.1103/PhysRevB.103.104505}.

\bibitem[{\citenamefont{Kaneko et~al.}(2019)\citenamefont{Kaneko, Shirakawa,
  Sorella, and Yunoki}}]{kaneko19}
\bibinfo{author}{\bibfnamefont{T.}~\bibnamefont{Kaneko}},
  \bibinfo{author}{\bibfnamefont{T.}~\bibnamefont{Shirakawa}},
  \bibinfo{author}{\bibfnamefont{S.}~\bibnamefont{Sorella}}, \bibnamefont{and}
  \bibinfo{author}{\bibfnamefont{S.}~\bibnamefont{Yunoki}},
  \bibinfo{journal}{Phys. Rev. Lett.} \textbf{\bibinfo{volume}{122}},
  \bibinfo{pages}{077002} (\bibinfo{year}{2019}),
  \urlprefix\url{https://link.aps.org/doi/10.1103/PhysRevLett.122.077002}.

\bibitem[{\citenamefont{Tindall et~al.}(2019)\citenamefont{Tindall, Bu\v{c}a,
  Coulthard, and Jaksch}}]{tindall19}
\bibinfo{author}{\bibfnamefont{J.}~\bibnamefont{Tindall}},
  \bibinfo{author}{\bibfnamefont{B.}~\bibnamefont{Bu\v{c}a}},
  \bibinfo{author}{\bibfnamefont{J.~R.} \bibnamefont{Coulthard}},
  \bibnamefont{and} \bibinfo{author}{\bibfnamefont{D.}~\bibnamefont{Jaksch}},
  \bibinfo{journal}{Phys. Rev. Lett.} \textbf{\bibinfo{volume}{123}},
  \bibinfo{pages}{030603} (\bibinfo{year}{2019}),
  \urlprefix\url{https://link.aps.org/doi/10.1103/PhysRevLett.123.030603}.

\bibitem[{\citenamefont{Ejima et~al.}(2020)\citenamefont{Ejima, Kaneko, Lange,
  Yunoki, and Fehske}}]{ejima20}
\bibinfo{author}{\bibfnamefont{S.}~\bibnamefont{Ejima}},
  \bibinfo{author}{\bibfnamefont{T.}~\bibnamefont{Kaneko}},
  \bibinfo{author}{\bibfnamefont{F.}~\bibnamefont{Lange}},
  \bibinfo{author}{\bibfnamefont{S.}~\bibnamefont{Yunoki}}, \bibnamefont{and}
  \bibinfo{author}{\bibfnamefont{H.}~\bibnamefont{Fehske}},
  \bibinfo{journal}{Phys. Rev. Res.} \textbf{\bibinfo{volume}{2}},
  \bibinfo{pages}{032008(R)} (\bibinfo{year}{2020}),
  \urlprefix\url{https://link.aps.org/doi/10.1103/PhysRevResearch.2.032008}.

\bibitem[{\citenamefont{Murakami et~al.}(2022)\citenamefont{Murakami,
  Takayoshi, Kaneko, Sun, Gole{\v z}, Millis, and Werner}}]{murakami22}
\bibinfo{author}{\bibfnamefont{Y.}~\bibnamefont{Murakami}},
  \bibinfo{author}{\bibfnamefont{S.}~\bibnamefont{Takayoshi}},
  \bibinfo{author}{\bibfnamefont{T.}~\bibnamefont{Kaneko}},
  \bibinfo{author}{\bibfnamefont{Z.}~\bibnamefont{Sun}},
  \bibinfo{author}{\bibfnamefont{D.}~\bibnamefont{Gole{\v z}}},
  \bibinfo{author}{\bibfnamefont{A.~J.} \bibnamefont{Millis}},
  \bibnamefont{and} \bibinfo{author}{\bibfnamefont{P.}~\bibnamefont{Werner}},
  \bibinfo{journal}{Communications Physics} \textbf{\bibinfo{volume}{5}},
  \bibinfo{pages}{23} (\bibinfo{year}{2022}),
  \urlprefix\url{https://doi.org/10.1038/s42005-021-00799-7}.

\bibitem[{\citenamefont{Werner et~al.}(2019)\citenamefont{Werner, Li,
  Gole\v{z}, and Eckstein}}]{werner19}
\bibinfo{author}{\bibfnamefont{P.}~\bibnamefont{Werner}},
  \bibinfo{author}{\bibfnamefont{J.}~\bibnamefont{Li}},
  \bibinfo{author}{\bibfnamefont{D.}~\bibnamefont{Gole\v{z}}},
  \bibnamefont{and} \bibinfo{author}{\bibfnamefont{M.}~\bibnamefont{Eckstein}},
  \bibinfo{journal}{Phys. Rev. B} \textbf{\bibinfo{volume}{100}},
  \bibinfo{pages}{155130} (\bibinfo{year}{2019}),
  \urlprefix\url{https://link.aps.org/doi/10.1103/PhysRevB.100.155130}.

\bibitem[{\citenamefont{Schlawin et~al.}(2019)\citenamefont{Schlawin,
  Cavalleri, and Jaksch}}]{schlawin19}
\bibinfo{author}{\bibfnamefont{F.}~\bibnamefont{Schlawin}},
  \bibinfo{author}{\bibfnamefont{A.}~\bibnamefont{Cavalleri}},
  \bibnamefont{and} \bibinfo{author}{\bibfnamefont{D.}~\bibnamefont{Jaksch}},
  \bibinfo{journal}{Phys. Rev. Lett.} \textbf{\bibinfo{volume}{122}},
  \bibinfo{pages}{133602} (\bibinfo{year}{2019}),
  \urlprefix\url{https://link.aps.org/doi/10.1103/PhysRevLett.122.133602}.

\bibitem[{\citenamefont{Li et~al.}(2023)\citenamefont{Li, M\"uller, Kim,
  L\"auchli, and Werner}}]{li23}
\bibinfo{author}{\bibfnamefont{J.}~\bibnamefont{Li}},
  \bibinfo{author}{\bibfnamefont{M.}~\bibnamefont{M\"uller}},
  \bibinfo{author}{\bibfnamefont{A.~J.} \bibnamefont{Kim}},
  \bibinfo{author}{\bibfnamefont{A.~M.} \bibnamefont{L\"auchli}},
  \bibnamefont{and} \bibinfo{author}{\bibfnamefont{P.}~\bibnamefont{Werner}},
  \bibinfo{journal}{Phys. Rev. B} \textbf{\bibinfo{volume}{107}},
  \bibinfo{pages}{205115} (\bibinfo{year}{2023}),
  \urlprefix\url{https://link.aps.org/doi/10.1103/PhysRevB.107.205115}.

\bibitem[{\citenamefont{Floquet}(1883)}]{floquet83}
\bibinfo{author}{\bibfnamefont{G.}~\bibnamefont{Floquet}},
  \bibinfo{journal}{Ann. Sci. \'{E}c. Norm. Sup\'{e}r.}
  \textbf{\bibinfo{volume}{12}}, \bibinfo{pages}{47} (\bibinfo{year}{1883}),
  \urlprefix\url{http://www.numdam.org/articles/10.24033/asens.220/}.

\bibitem[{\citenamefont{Shirley}(1965)}]{shirley65}
\bibinfo{author}{\bibfnamefont{J.~H.} \bibnamefont{Shirley}},
  \bibinfo{journal}{Phys. Rev.} \textbf{\bibinfo{volume}{138}},
  \bibinfo{pages}{B979} (\bibinfo{year}{1965}),
  \urlprefix\url{https://link.aps.org/doi/10.1103/PhysRev.138.B979}.

\bibitem[{\citenamefont{Ivanov et~al.}(2021)\citenamefont{Ivanov, Mote, Ernst,
  Equbal, and Madhu}}]{ivanov21}
\bibinfo{author}{\bibfnamefont{K.~L.} \bibnamefont{Ivanov}},
  \bibinfo{author}{\bibfnamefont{K.~R.} \bibnamefont{Mote}},
  \bibinfo{author}{\bibfnamefont{M.}~\bibnamefont{Ernst}},
  \bibinfo{author}{\bibfnamefont{A.}~\bibnamefont{Equbal}}, \bibnamefont{and}
  \bibinfo{author}{\bibfnamefont{P.~K.} \bibnamefont{Madhu}},
  \bibinfo{journal}{Progress in Nuclear Magnetic Resonance Spectroscopy}
  \textbf{\bibinfo{volume}{126-127}}, \bibinfo{pages}{17}
  (\bibinfo{year}{2021}), ISSN \bibinfo{issn}{0079-6565},
  \urlprefix\url{https://www.sciencedirect.com/science/article/pii/S0079656521000169}.

\bibitem[{\citenamefont{Ozawa and Price}(2019)}]{ozawa19}
\bibinfo{author}{\bibfnamefont{T.}~\bibnamefont{Ozawa}} \bibnamefont{and}
  \bibinfo{author}{\bibfnamefont{H.~M.} \bibnamefont{Price}},
  \bibinfo{journal}{Nature Reviews Physics} \textbf{\bibinfo{volume}{1}},
  \bibinfo{pages}{349} (\bibinfo{year}{2019}),
  \urlprefix\url{https://doi.org/10.1038/s42254-019-0045-3}.

\bibitem[{\citenamefont{Price et~al.}(2015)\citenamefont{Price, Zilberberg,
  Ozawa, Carusotto, and Goldman}}]{price15}
\bibinfo{author}{\bibfnamefont{H.~M.} \bibnamefont{Price}},
  \bibinfo{author}{\bibfnamefont{O.}~\bibnamefont{Zilberberg}},
  \bibinfo{author}{\bibfnamefont{T.}~\bibnamefont{Ozawa}},
  \bibinfo{author}{\bibfnamefont{I.}~\bibnamefont{Carusotto}},
  \bibnamefont{and} \bibinfo{author}{\bibfnamefont{N.}~\bibnamefont{Goldman}},
  \bibinfo{journal}{Phys. Rev. Lett.} \textbf{\bibinfo{volume}{115}},
  \bibinfo{pages}{195303} (\bibinfo{year}{2015}),
  \urlprefix\url{https://link.aps.org/doi/10.1103/PhysRevLett.115.195303}.

\bibitem[{\citenamefont{Price et~al.}(2017)\citenamefont{Price, Ozawa, and
  Goldman}}]{price17}
\bibinfo{author}{\bibfnamefont{H.~M.} \bibnamefont{Price}},
  \bibinfo{author}{\bibfnamefont{T.}~\bibnamefont{Ozawa}}, \bibnamefont{and}
  \bibinfo{author}{\bibfnamefont{N.}~\bibnamefont{Goldman}},
  \bibinfo{journal}{Phys. Rev. A} \textbf{\bibinfo{volume}{95}},
  \bibinfo{pages}{023607} (\bibinfo{year}{2017}),
  \urlprefix\url{https://link.aps.org/doi/10.1103/PhysRevA.95.023607}.

\bibitem[{\citenamefont{Lustig et~al.}(2019)\citenamefont{Lustig, Weimann,
  Plotnik, Lumer, Bandres, Szameit, and Segev}}]{lustig19}
\bibinfo{author}{\bibfnamefont{E.}~\bibnamefont{Lustig}},
  \bibinfo{author}{\bibfnamefont{S.}~\bibnamefont{Weimann}},
  \bibinfo{author}{\bibfnamefont{Y.}~\bibnamefont{Plotnik}},
  \bibinfo{author}{\bibfnamefont{Y.}~\bibnamefont{Lumer}},
  \bibinfo{author}{\bibfnamefont{M.~A.} \bibnamefont{Bandres}},
  \bibinfo{author}{\bibfnamefont{A.}~\bibnamefont{Szameit}}, \bibnamefont{and}
  \bibinfo{author}{\bibfnamefont{M.}~\bibnamefont{Segev}},
  \bibinfo{journal}{Nature} \textbf{\bibinfo{volume}{567}},
  \bibinfo{pages}{356} (\bibinfo{year}{2019}),
  \urlprefix\url{https://doi.org/10.1038/s41586-019-0943-7}.

\bibitem[{\citenamefont{Verstraete et~al.}(2009)\citenamefont{Verstraete, Wolf,
  and Ignacio~Cirac}}]{verstraete09}
\bibinfo{author}{\bibfnamefont{F.}~\bibnamefont{Verstraete}},
  \bibinfo{author}{\bibfnamefont{M.~M.} \bibnamefont{Wolf}}, \bibnamefont{and}
  \bibinfo{author}{\bibfnamefont{J.}~\bibnamefont{Ignacio~Cirac}},
  \bibinfo{journal}{Nature Physics} \textbf{\bibinfo{volume}{5}},
  \bibinfo{pages}{633} (\bibinfo{year}{2009}),
  \urlprefix\url{https://doi.org/10.1038/nphys1342}.

\bibitem[{\citenamefont{Sieberer et~al.}(2016)\citenamefont{Sieberer, Buchhold,
  and Diehl}}]{sieberer16}
\bibinfo{author}{\bibfnamefont{L.~M.} \bibnamefont{Sieberer}},
  \bibinfo{author}{\bibfnamefont{M.}~\bibnamefont{Buchhold}}, \bibnamefont{and}
  \bibinfo{author}{\bibfnamefont{S.}~\bibnamefont{Diehl}},
  \bibinfo{journal}{Reports on Progress in Physics}
  \textbf{\bibinfo{volume}{79}}, \bibinfo{pages}{096001}
  (\bibinfo{year}{2016}),
  \urlprefix\url{https://doi.org/10.1088/0034-4885/79/9/096001}.

\bibitem[{\citenamefont{Manzano}(2020)}]{manzano20}
\bibinfo{author}{\bibfnamefont{D.}~\bibnamefont{Manzano}},
  \bibinfo{journal}{AIP Advances} \textbf{\bibinfo{volume}{10}},
  \bibinfo{pages}{025106} (\bibinfo{year}{2020}),
  \urlprefix\url{https://doi.org/10.1063/1.5115323}.

\bibitem[{\citenamefont{S\'anchez Mu\~noz et~al.}(2019)\citenamefont{S\'anchez
  Mu\~noz, Bu\v{c}a, Tindall, Gonz\'alez-Tudela, Jaksch, and
  Porras}}]{sanchez19}
\bibinfo{author}{\bibfnamefont{C.}~\bibnamefont{S\'anchez Mu\~noz}},
  \bibinfo{author}{\bibfnamefont{B.}~\bibnamefont{Bu\v{c}a}},
  \bibinfo{author}{\bibfnamefont{J.}~\bibnamefont{Tindall}},
  \bibinfo{author}{\bibfnamefont{A.}~\bibnamefont{Gonz\'alez-Tudela}},
  \bibinfo{author}{\bibfnamefont{D.}~\bibnamefont{Jaksch}}, \bibnamefont{and}
  \bibinfo{author}{\bibfnamefont{D.}~\bibnamefont{Porras}},
  \bibinfo{journal}{Phys. Rev. A} \textbf{\bibinfo{volume}{100}},
  \bibinfo{pages}{042113} (\bibinfo{year}{2019}),
  \urlprefix\url{https://link.aps.org/doi/10.1103/PhysRevA.100.042113}.

\bibitem[{\citenamefont{Tindall et~al.}(2020)\citenamefont{Tindall,
  Sánchez~Muñoz, Buča, and Jaksch}}]{tindall20}
\bibinfo{author}{\bibfnamefont{J.}~\bibnamefont{Tindall}},
  \bibinfo{author}{\bibfnamefont{C.}~\bibnamefont{Sánchez~Muñoz}},
  \bibinfo{author}{\bibfnamefont{B.}~\bibnamefont{Buča}}, \bibnamefont{and}
  \bibinfo{author}{\bibfnamefont{D.}~\bibnamefont{Jaksch}},
  \bibinfo{journal}{New Journal of Physics} \textbf{\bibinfo{volume}{22}},
  \bibinfo{pages}{013026} (\bibinfo{year}{2020}),
  \urlprefix\url{https://dx.doi.org/10.1088/1367-2630/ab60f5}.

\bibitem[{\citenamefont{Nagy et~al.}(2011)\citenamefont{Nagy, Szirmai, and
  Domokos}}]{nagy11}
\bibinfo{author}{\bibfnamefont{D.}~\bibnamefont{Nagy}},
  \bibinfo{author}{\bibfnamefont{G.}~\bibnamefont{Szirmai}}, \bibnamefont{and}
  \bibinfo{author}{\bibfnamefont{P.}~\bibnamefont{Domokos}},
  \bibinfo{journal}{Phys. Rev. A} \textbf{\bibinfo{volume}{84}},
  \bibinfo{pages}{043637} (\bibinfo{year}{2011}),
  \urlprefix\url{https://link.aps.org/doi/10.1103/PhysRevA.84.043637}.

\bibitem[{\citenamefont{Brennecke et~al.}(2013)\citenamefont{Brennecke, Mottl,
  Baumann, Landig, Donner, and Esslinger}}]{brennecke13}
\bibinfo{author}{\bibfnamefont{F.}~\bibnamefont{Brennecke}},
  \bibinfo{author}{\bibfnamefont{R.}~\bibnamefont{Mottl}},
  \bibinfo{author}{\bibfnamefont{K.}~\bibnamefont{Baumann}},
  \bibinfo{author}{\bibfnamefont{R.}~\bibnamefont{Landig}},
  \bibinfo{author}{\bibfnamefont{T.}~\bibnamefont{Donner}}, \bibnamefont{and}
  \bibinfo{author}{\bibfnamefont{T.}~\bibnamefont{Esslinger}},
  \bibinfo{journal}{Proceedings of the National Academy of Sciences}
  \textbf{\bibinfo{volume}{110}}, \bibinfo{pages}{11763}
  (\bibinfo{year}{2013}),
  \urlprefix\url{http://www.pnas.org/content/110/29/11763}.

\bibitem[{\citenamefont{Torre et~al.}(2013)\citenamefont{Torre, Diehl, Lukin,
  Sachdev, and Strack}}]{torre13}
\bibinfo{author}{\bibfnamefont{E.~G.~D.} \bibnamefont{Torre}},
  \bibinfo{author}{\bibfnamefont{S.}~\bibnamefont{Diehl}},
  \bibinfo{author}{\bibfnamefont{M.~D.} \bibnamefont{Lukin}},
  \bibinfo{author}{\bibfnamefont{S.}~\bibnamefont{Sachdev}}, \bibnamefont{and}
  \bibinfo{author}{\bibfnamefont{P.}~\bibnamefont{Strack}},
  \bibinfo{journal}{Phys. Rev. A} \textbf{\bibinfo{volume}{87}},
  \bibinfo{pages}{023831} (\bibinfo{year}{2013}),
  \urlprefix\url{https://link.aps.org/doi/10.1103/PhysRevA.87.023831}.

\bibitem[{\citenamefont{Soriente et~al.}(2018)\citenamefont{Soriente, Donner,
  Chitra, and Zilberberg}}]{soriente18}
\bibinfo{author}{\bibfnamefont{M.}~\bibnamefont{Soriente}},
  \bibinfo{author}{\bibfnamefont{T.}~\bibnamefont{Donner}},
  \bibinfo{author}{\bibfnamefont{R.}~\bibnamefont{Chitra}}, \bibnamefont{and}
  \bibinfo{author}{\bibfnamefont{O.}~\bibnamefont{Zilberberg}},
  \bibinfo{journal}{Phys. Rev. Lett.} \textbf{\bibinfo{volume}{120}},
  \bibinfo{pages}{183603} (\bibinfo{year}{2018}),
  \urlprefix\url{https://link.aps.org/doi/10.1103/PhysRevLett.120.183603}.

\bibitem[{\citenamefont{Soriente et~al.}(2020)\citenamefont{Soriente, Chitra,
  and Zilberberg}}]{soriente20}
\bibinfo{author}{\bibfnamefont{M.}~\bibnamefont{Soriente}},
  \bibinfo{author}{\bibfnamefont{R.}~\bibnamefont{Chitra}}, \bibnamefont{and}
  \bibinfo{author}{\bibfnamefont{O.}~\bibnamefont{Zilberberg}},
  \bibinfo{journal}{Phys. Rev. A} \textbf{\bibinfo{volume}{101}},
  \bibinfo{pages}{023823} (\bibinfo{year}{2020}),
  \urlprefix\url{https://link.aps.org/doi/10.1103/PhysRevA.101.023823}.

\bibitem[{\citenamefont{Ferri et~al.}(2021)\citenamefont{Ferri, Rosa-Medina,
  Finger, Dogra, Soriente, Zilberberg, Donner, and Esslinger}}]{ferri21}
\bibinfo{author}{\bibfnamefont{F.}~\bibnamefont{Ferri}},
  \bibinfo{author}{\bibfnamefont{R.}~\bibnamefont{Rosa-Medina}},
  \bibinfo{author}{\bibfnamefont{F.}~\bibnamefont{Finger}},
  \bibinfo{author}{\bibfnamefont{N.}~\bibnamefont{Dogra}},
  \bibinfo{author}{\bibfnamefont{M.}~\bibnamefont{Soriente}},
  \bibinfo{author}{\bibfnamefont{O.}~\bibnamefont{Zilberberg}},
  \bibinfo{author}{\bibfnamefont{T.}~\bibnamefont{Donner}}, \bibnamefont{and}
  \bibinfo{author}{\bibfnamefont{T.}~\bibnamefont{Esslinger}},
  \bibinfo{journal}{Phys. Rev. X} \textbf{\bibinfo{volume}{11}},
  \bibinfo{pages}{041046} (\bibinfo{year}{2021}),
  \urlprefix\url{https://link.aps.org/doi/10.1103/PhysRevX.11.041046}.

\bibitem[{\citenamefont{Lin et~al.}(2022)\citenamefont{Lin, Rosa-Medina, Ferri,
  Finger, Kroeger, Donner, Esslinger, and Chitra}}]{lin22}
\bibinfo{author}{\bibfnamefont{R.}~\bibnamefont{Lin}},
  \bibinfo{author}{\bibfnamefont{R.}~\bibnamefont{Rosa-Medina}},
  \bibinfo{author}{\bibfnamefont{F.}~\bibnamefont{Ferri}},
  \bibinfo{author}{\bibfnamefont{F.}~\bibnamefont{Finger}},
  \bibinfo{author}{\bibfnamefont{K.}~\bibnamefont{Kroeger}},
  \bibinfo{author}{\bibfnamefont{T.}~\bibnamefont{Donner}},
  \bibinfo{author}{\bibfnamefont{T.}~\bibnamefont{Esslinger}},
  \bibnamefont{and} \bibinfo{author}{\bibfnamefont{R.}~\bibnamefont{Chitra}},
  \bibinfo{journal}{Phys. Rev. Lett.} \textbf{\bibinfo{volume}{128}},
  \bibinfo{pages}{153601} (\bibinfo{year}{2022}),
  \urlprefix\url{https://link.aps.org/doi/10.1103/PhysRevLett.128.153601}.

\bibitem[{\citenamefont{Rosa-Medina et~al.}(2022)\citenamefont{Rosa-Medina,
  Ferri, Finger, Dogra, Kroeger, Lin, Chitra, Donner, and
  Esslinger}}]{rosamedina22}
\bibinfo{author}{\bibfnamefont{R.}~\bibnamefont{Rosa-Medina}},
  \bibinfo{author}{\bibfnamefont{F.}~\bibnamefont{Ferri}},
  \bibinfo{author}{\bibfnamefont{F.}~\bibnamefont{Finger}},
  \bibinfo{author}{\bibfnamefont{N.}~\bibnamefont{Dogra}},
  \bibinfo{author}{\bibfnamefont{K.}~\bibnamefont{Kroeger}},
  \bibinfo{author}{\bibfnamefont{R.}~\bibnamefont{Lin}},
  \bibinfo{author}{\bibfnamefont{R.}~\bibnamefont{Chitra}},
  \bibinfo{author}{\bibfnamefont{T.}~\bibnamefont{Donner}}, \bibnamefont{and}
  \bibinfo{author}{\bibfnamefont{T.}~\bibnamefont{Esslinger}},
  \bibinfo{journal}{Phys. Rev. Lett.} \textbf{\bibinfo{volume}{128}},
  \bibinfo{pages}{143602} (\bibinfo{year}{2022}),
  \urlprefix\url{https://link.aps.org/doi/10.1103/PhysRevLett.128.143602}.

\bibitem[{\citenamefont{Keldysh}(1964)}]{keldysh64}
\bibinfo{author}{\bibfnamefont{L.~V.} \bibnamefont{Keldysh}},
  \bibinfo{journal}{Zh. Eksp. Teor. Fiz.} \textbf{\bibinfo{volume}{47}},
  \bibinfo{pages}{1515–1527} (\bibinfo{year}{1964}),
  \urlprefix\url{http://jetp.ras.ru/cgi-bin/e/index/r/47/4/p1515?a=list}.

\bibitem[{\citenamefont{Altland and Simons}(2010)}]{altland10}
\bibinfo{author}{\bibfnamefont{A.}~\bibnamefont{Altland}} \bibnamefont{and}
  \bibinfo{author}{\bibfnamefont{B.~D.} \bibnamefont{Simons}},
  \emph{\bibinfo{title}{Condensed Matter Field Theory}}
  (\bibinfo{publisher}{Cambridge University Press}, \bibinfo{year}{2010}),
  \bibinfo{edition}{2nd} ed.

\bibitem[{\citenamefont{Aoki et~al.}(2014)\citenamefont{Aoki, Tsuji, Eckstein,
  Kollar, Oka, and Werner}}]{aoki14}
\bibinfo{author}{\bibfnamefont{H.}~\bibnamefont{Aoki}},
  \bibinfo{author}{\bibfnamefont{N.}~\bibnamefont{Tsuji}},
  \bibinfo{author}{\bibfnamefont{M.}~\bibnamefont{Eckstein}},
  \bibinfo{author}{\bibfnamefont{M.}~\bibnamefont{Kollar}},
  \bibinfo{author}{\bibfnamefont{T.}~\bibnamefont{Oka}}, \bibnamefont{and}
  \bibinfo{author}{\bibfnamefont{P.}~\bibnamefont{Werner}},
  \bibinfo{journal}{Rev. Mod. Phys.} \textbf{\bibinfo{volume}{86}},
  \bibinfo{pages}{779} (\bibinfo{year}{2014}),
  \urlprefix\url{https://link.aps.org/doi/10.1103/RevModPhys.86.779}.

\bibitem[{\citenamefont{Ramires and Sigrist}(2016)}]{ramires16}
\bibinfo{author}{\bibfnamefont{A.}~\bibnamefont{Ramires}} \bibnamefont{and}
  \bibinfo{author}{\bibfnamefont{M.}~\bibnamefont{Sigrist}},
  \bibinfo{journal}{Phys. Rev. B} \textbf{\bibinfo{volume}{94}},
  \bibinfo{pages}{104501} (\bibinfo{year}{2016}),
  \urlprefix\url{https://link.aps.org/doi/10.1103/PhysRevB.94.104501}.

\bibitem[{\citenamefont{Ramires et~al.}(2018)\citenamefont{Ramires, Agterberg,
  and Sigrist}}]{ramires18}
\bibinfo{author}{\bibfnamefont{A.}~\bibnamefont{Ramires}},
  \bibinfo{author}{\bibfnamefont{D.~F.} \bibnamefont{Agterberg}},
  \bibnamefont{and} \bibinfo{author}{\bibfnamefont{M.}~\bibnamefont{Sigrist}},
  \bibinfo{journal}{Phys. Rev. B} \textbf{\bibinfo{volume}{98}},
  \bibinfo{pages}{024501} (\bibinfo{year}{2018}),
  \urlprefix\url{https://link.aps.org/doi/10.1103/PhysRevB.98.024501}.

\bibitem[{foo()}]{footnote1}
\bibinfo{note}{The fitness measure has been given explicitly in Nambu space in
  Eq.~\eqref{eq:commutativity}, but instead in terms of the Hamiltonian and gap
  matrices respectively in particle-particle and particle-hole spaces in
  Refs.~\cite{ramires16,ramires18}.}

\bibitem[{\citenamefont{Tsuji et~al.}(2008)\citenamefont{Tsuji, Oka, and
  Aoki}}]{tsuji08}
\bibinfo{author}{\bibfnamefont{N.}~\bibnamefont{Tsuji}},
  \bibinfo{author}{\bibfnamefont{T.}~\bibnamefont{Oka}}, \bibnamefont{and}
  \bibinfo{author}{\bibfnamefont{H.}~\bibnamefont{Aoki}},
  \bibinfo{journal}{Phys. Rev. B} \textbf{\bibinfo{volume}{78}},
  \bibinfo{pages}{235124} (\bibinfo{year}{2008}),
  \urlprefix\url{https://link.aps.org/doi/10.1103/PhysRevB.78.235124}.

\bibitem[{sup()}]{supmat}
\bibinfo{note}{See Supplemental Material for an introduction to the
  Shirley-Floquet formalism, a derivation for the Green's function in this
  formalism, a discussion to higher harmonics of the oscillation, and the
  application to dispersive superconductors.}

\bibitem[{\citenamefont{Shi et~al.}(2023)\citenamefont{Shi, Matsyshyn, Song,
  and Villadiego}}]{shi23}
\bibinfo{author}{\bibfnamefont{L.-K.} \bibnamefont{Shi}},
  \bibinfo{author}{\bibfnamefont{O.}~\bibnamefont{Matsyshyn}},
  \bibinfo{author}{\bibfnamefont{J.~C.~W.} \bibnamefont{Song}},
  \bibnamefont{and} \bibinfo{author}{\bibfnamefont{I.~S.}
  \bibnamefont{Villadiego}}, \emph{\bibinfo{title}{The floquet fermi liquid}}
  (\bibinfo{year}{2023}), \eprint{2309.03268}.

\bibitem[{\citenamefont{Yang et~al.}(2021)\citenamefont{Yang, Yang, and
  Liu}}]{yang21}
\bibinfo{author}{\bibfnamefont{Q.}~\bibnamefont{Yang}},
  \bibinfo{author}{\bibfnamefont{Z.}~\bibnamefont{Yang}}, \bibnamefont{and}
  \bibinfo{author}{\bibfnamefont{D.~E.} \bibnamefont{Liu}},
  \bibinfo{journal}{Phys. Rev. B} \textbf{\bibinfo{volume}{104}},
  \bibinfo{pages}{014512} (\bibinfo{year}{2021}),
  \urlprefix\url{https://link.aps.org/doi/10.1103/PhysRevB.104.014512}.

\bibitem[{\citenamefont{Uhrig et~al.}(2019)\citenamefont{Uhrig, Kalthoff, and
  Freericks}}]{uhrig19}
\bibinfo{author}{\bibfnamefont{G.~S.} \bibnamefont{Uhrig}},
  \bibinfo{author}{\bibfnamefont{M.~H.} \bibnamefont{Kalthoff}},
  \bibnamefont{and} \bibinfo{author}{\bibfnamefont{J.~K.}
  \bibnamefont{Freericks}}, \bibinfo{journal}{Phys. Rev. Lett.}
  \textbf{\bibinfo{volume}{122}}, \bibinfo{pages}{130604}
  (\bibinfo{year}{2019}),
  \urlprefix\url{https://link.aps.org/doi/10.1103/PhysRevLett.122.130604}.

\bibitem[{\citenamefont{Chubukov et~al.}(2020)\citenamefont{Chubukov, Abanov,
  Esterlis, and Kivelson}}]{chubukov20}
\bibinfo{author}{\bibfnamefont{A.~V.} \bibnamefont{Chubukov}},
  \bibinfo{author}{\bibfnamefont{A.}~\bibnamefont{Abanov}},
  \bibinfo{author}{\bibfnamefont{I.}~\bibnamefont{Esterlis}}, \bibnamefont{and}
  \bibinfo{author}{\bibfnamefont{S.~A.} \bibnamefont{Kivelson}},
  \bibinfo{journal}{Annals of Physics} \textbf{\bibinfo{volume}{417}},
  \bibinfo{pages}{168190} (\bibinfo{year}{2020}), ISSN
  \bibinfo{issn}{0003-4916}, \bibinfo{note}{eliashberg theory at 60:
  Strong-coupling superconductivity and beyond},
  \urlprefix\url{https://www.sciencedirect.com/science/article/pii/S0003491620301238}.

\bibitem[{\citenamefont{Marsiglio}(2020)}]{marsiglio20}
\bibinfo{author}{\bibfnamefont{F.}~\bibnamefont{Marsiglio}},
  \bibinfo{journal}{Annals of Physics} \textbf{\bibinfo{volume}{417}},
  \bibinfo{pages}{168102} (\bibinfo{year}{2020}), ISSN
  \bibinfo{issn}{0003-4916}, \bibinfo{note}{eliashberg theory at 60:
  Strong-coupling superconductivity and beyond},
  \urlprefix\url{https://www.sciencedirect.com/science/article/pii/S000349162030035X}.

\bibitem[{\citenamefont{You and Vishwanath}(2019)}]{you19}
\bibinfo{author}{\bibfnamefont{Y.-Z.} \bibnamefont{You}} \bibnamefont{and}
  \bibinfo{author}{\bibfnamefont{A.}~\bibnamefont{Vishwanath}},
  \bibinfo{journal}{npj Quantum Materials} \textbf{\bibinfo{volume}{4}},
  \bibinfo{pages}{16} (\bibinfo{year}{2019}),
  \urlprefix\url{https://doi.org/10.1038/s41535-019-0153-4}.

\bibitem[{\citenamefont{Liu et~al.}(2022)\citenamefont{Liu, Bourges, Sidis,
  Xie, He, Bourdarot, Danilkin, Ghosh, Ghosh, Ma et~al.}}]{liu22}
\bibinfo{author}{\bibfnamefont{C.}~\bibnamefont{Liu}},
  \bibinfo{author}{\bibfnamefont{P.}~\bibnamefont{Bourges}},
  \bibinfo{author}{\bibfnamefont{Y.}~\bibnamefont{Sidis}},
  \bibinfo{author}{\bibfnamefont{T.}~\bibnamefont{Xie}},
  \bibinfo{author}{\bibfnamefont{G.}~\bibnamefont{He}},
  \bibinfo{author}{\bibfnamefont{F.}~\bibnamefont{Bourdarot}},
  \bibinfo{author}{\bibfnamefont{S.}~\bibnamefont{Danilkin}},
  \bibinfo{author}{\bibfnamefont{H.}~\bibnamefont{Ghosh}},
  \bibinfo{author}{\bibfnamefont{S.}~\bibnamefont{Ghosh}},
  \bibinfo{author}{\bibfnamefont{X.}~\bibnamefont{Ma}}, \bibnamefont{et~al.},
  \bibinfo{journal}{Phys. Rev. Lett.} \textbf{\bibinfo{volume}{128}},
  \bibinfo{pages}{137003} (\bibinfo{year}{2022}),
  \urlprefix\url{https://link.aps.org/doi/10.1103/PhysRevLett.128.137003}.

\bibitem[{\citenamefont{Chakraborty and Piazza}(2021)}]{chakraborty21}
\bibinfo{author}{\bibfnamefont{A.}~\bibnamefont{Chakraborty}} \bibnamefont{and}
  \bibinfo{author}{\bibfnamefont{F.}~\bibnamefont{Piazza}},
  \bibinfo{journal}{Phys. Rev. Lett.} \textbf{\bibinfo{volume}{127}},
  \bibinfo{pages}{177002} (\bibinfo{year}{2021}),
  \urlprefix\url{https://link.aps.org/doi/10.1103/PhysRevLett.127.177002}.

\bibitem[{\citenamefont{Kelly et~al.}(2022)\citenamefont{Kelly, Thompson, Rey,
  and Marino}}]{kelly22}
\bibinfo{author}{\bibfnamefont{S.~P.} \bibnamefont{Kelly}},
  \bibinfo{author}{\bibfnamefont{J.~K.} \bibnamefont{Thompson}},
  \bibinfo{author}{\bibfnamefont{A.~M.} \bibnamefont{Rey}}, \bibnamefont{and}
  \bibinfo{author}{\bibfnamefont{J.}~\bibnamefont{Marino}},
  \bibinfo{journal}{Phys. Rev. Res.} \textbf{\bibinfo{volume}{4}},
  \bibinfo{pages}{L042032} (\bibinfo{year}{2022}),
  \urlprefix\url{https://link.aps.org/doi/10.1103/PhysRevResearch.4.L042032}.

\bibitem[{\citenamefont{Hart et~al.}(2019)\citenamefont{Hart, Goldstein,
  Chamon, and Castelnovo}}]{hart19}
\bibinfo{author}{\bibfnamefont{O.}~\bibnamefont{Hart}},
  \bibinfo{author}{\bibfnamefont{G.}~\bibnamefont{Goldstein}},
  \bibinfo{author}{\bibfnamefont{C.}~\bibnamefont{Chamon}}, \bibnamefont{and}
  \bibinfo{author}{\bibfnamefont{C.}~\bibnamefont{Castelnovo}},
  \bibinfo{journal}{Phys. Rev. B} \textbf{\bibinfo{volume}{100}},
  \bibinfo{pages}{060508(R)} (\bibinfo{year}{2019}),
  \urlprefix\url{https://link.aps.org/doi/10.1103/PhysRevB.100.060508}.

\bibitem[{\citenamefont{Eliashberg}(1971)}]{eliashberg71}
\bibinfo{author}{\bibfnamefont{G.~M.} \bibnamefont{Eliashberg}},
  \bibinfo{journal}{Zh. Eksp. Teor. Fiz.} \textbf{\bibinfo{volume}{3}},
  \bibinfo{pages}{1254} (\bibinfo{year}{1971}),
  \urlprefix\url{http://www.jetp.ras.ru/cgi-bin/e/index/r/61/3/p1254?a=list}.

\bibitem[{\citenamefont{Ivlev et~al.}(1973)\citenamefont{Ivlev, Lisitsyn, and
  Eliashberg}}]{ivlev73}
\bibinfo{author}{\bibfnamefont{B.~I.} \bibnamefont{Ivlev}},
  \bibinfo{author}{\bibfnamefont{S.~G.} \bibnamefont{Lisitsyn}},
  \bibnamefont{and} \bibinfo{author}{\bibfnamefont{G.~M.}
  \bibnamefont{Eliashberg}}, \bibinfo{journal}{Journal of Low Temperature
  Physics} \textbf{\bibinfo{volume}{10}}, \bibinfo{pages}{449}
  (\bibinfo{year}{1973}), \urlprefix\url{https://doi.org/10.1007/BF00654920}.

\bibitem[{\citenamefont{Kennes et~al.}(2021)\citenamefont{Kennes, Claassen,
  Xian, Georges, Millis, Hone, Dean, Basov, Pasupathy, and Rubio}}]{kennes21}
\bibinfo{author}{\bibfnamefont{D.~M.} \bibnamefont{Kennes}},
  \bibinfo{author}{\bibfnamefont{M.}~\bibnamefont{Claassen}},
  \bibinfo{author}{\bibfnamefont{L.}~\bibnamefont{Xian}},
  \bibinfo{author}{\bibfnamefont{A.}~\bibnamefont{Georges}},
  \bibinfo{author}{\bibfnamefont{A.~J.} \bibnamefont{Millis}},
  \bibinfo{author}{\bibfnamefont{J.}~\bibnamefont{Hone}},
  \bibinfo{author}{\bibfnamefont{C.~R.} \bibnamefont{Dean}},
  \bibinfo{author}{\bibfnamefont{D.~N.} \bibnamefont{Basov}},
  \bibinfo{author}{\bibfnamefont{A.~N.} \bibnamefont{Pasupathy}},
  \bibnamefont{and} \bibinfo{author}{\bibfnamefont{A.}~\bibnamefont{Rubio}},
  \bibinfo{journal}{Nature Physics} \textbf{\bibinfo{volume}{17}},
  \bibinfo{pages}{155} (\bibinfo{year}{2021}),
  \urlprefix\url{https://doi.org/10.1038/s41567-020-01154-3}.

\bibitem[{\citenamefont{Ramires}(2022)}]{ramires22}
\bibinfo{author}{\bibfnamefont{A.}~\bibnamefont{Ramires}},
  \bibinfo{journal}{Nature} \textbf{\bibinfo{volume}{608}},
  \bibinfo{pages}{474} (\bibinfo{year}{2022}),
  \urlprefix\url{https://www.nature.com/articles/d41586-022-02108-w}.

\bibitem[{\citenamefont{Lai et~al.}(2018)\citenamefont{Lai, Grefe, Paschen, and
  Si}}]{lai18}
\bibinfo{author}{\bibfnamefont{H.-H.} \bibnamefont{Lai}},
  \bibinfo{author}{\bibfnamefont{S.~E.} \bibnamefont{Grefe}},
  \bibinfo{author}{\bibfnamefont{S.}~\bibnamefont{Paschen}}, \bibnamefont{and}
  \bibinfo{author}{\bibfnamefont{Q.}~\bibnamefont{Si}},
  \bibinfo{journal}{Proceedings of the National Academy of Sciences}
  \textbf{\bibinfo{volume}{115}}, \bibinfo{pages}{93} (\bibinfo{year}{2018}),
  \eprint{https://www.pnas.org/doi/pdf/10.1073/pnas.1715851115},
  \urlprefix\url{https://www.pnas.org/doi/abs/10.1073/pnas.1715851115}.

\bibitem[{\citenamefont{Regnault et~al.}(2022)\citenamefont{Regnault, Xu, Li,
  Ma, Jovanovic, Yazdani, Parkin, Felser, Schoop, Ong et~al.}}]{regnault22}
\bibinfo{author}{\bibfnamefont{N.}~\bibnamefont{Regnault}},
  \bibinfo{author}{\bibfnamefont{Y.}~\bibnamefont{Xu}},
  \bibinfo{author}{\bibfnamefont{M.-R.} \bibnamefont{Li}},
  \bibinfo{author}{\bibfnamefont{D.-S.} \bibnamefont{Ma}},
  \bibinfo{author}{\bibfnamefont{M.}~\bibnamefont{Jovanovic}},
  \bibinfo{author}{\bibfnamefont{A.}~\bibnamefont{Yazdani}},
  \bibinfo{author}{\bibfnamefont{S.~S.~P.} \bibnamefont{Parkin}},
  \bibinfo{author}{\bibfnamefont{C.}~\bibnamefont{Felser}},
  \bibinfo{author}{\bibfnamefont{L.~M.} \bibnamefont{Schoop}},
  \bibinfo{author}{\bibfnamefont{N.~P.} \bibnamefont{Ong}},
  \bibnamefont{et~al.}, \bibinfo{journal}{Nature}
  \textbf{\bibinfo{volume}{603}}, \bibinfo{pages}{824} (\bibinfo{year}{2022}),
  \urlprefix\url{https://doi.org/10.1038/s41586-022-04519-1}.

\bibitem[{\citenamefont{Wang et~al.}(2013)\citenamefont{Wang, Steinberg,
  Jarillo-Herrero, and Gedik}}]{wang13}
\bibinfo{author}{\bibfnamefont{Y.~H.} \bibnamefont{Wang}},
  \bibinfo{author}{\bibfnamefont{H.}~\bibnamefont{Steinberg}},
  \bibinfo{author}{\bibfnamefont{P.}~\bibnamefont{Jarillo-Herrero}},
  \bibnamefont{and} \bibinfo{author}{\bibfnamefont{N.}~\bibnamefont{Gedik}},
  \bibinfo{journal}{Science} \textbf{\bibinfo{volume}{342}},
  \bibinfo{pages}{453} (\bibinfo{year}{2013}),
  \eprint{https://www.science.org/doi/pdf/10.1126/science.1239834},
  \urlprefix\url{https://www.science.org/doi/abs/10.1126/science.1239834}.

\bibitem[{\citenamefont{McIver et~al.}(2020)\citenamefont{McIver, Schulte,
  Stein, Matsuyama, Jotzu, Meier, and Cavalleri}}]{mciver20}
\bibinfo{author}{\bibfnamefont{J.~W.} \bibnamefont{McIver}},
  \bibinfo{author}{\bibfnamefont{B.}~\bibnamefont{Schulte}},
  \bibinfo{author}{\bibfnamefont{F.~U.} \bibnamefont{Stein}},
  \bibinfo{author}{\bibfnamefont{T.}~\bibnamefont{Matsuyama}},
  \bibinfo{author}{\bibfnamefont{G.}~\bibnamefont{Jotzu}},
  \bibinfo{author}{\bibfnamefont{G.}~\bibnamefont{Meier}}, \bibnamefont{and}
  \bibinfo{author}{\bibfnamefont{A.}~\bibnamefont{Cavalleri}},
  \bibinfo{journal}{Nature Physics} \textbf{\bibinfo{volume}{16}},
  \bibinfo{pages}{38} (\bibinfo{year}{2020}),
  \urlprefix\url{https://doi.org/10.1038/s41567-019-0698-y}.

\end{thebibliography}


\begin{thebibliography}{4}
\expandafter\ifx\csname natexlab\endcsname\relax\def\natexlab#1{#1}\fi
\expandafter\ifx\csname bibnamefont\endcsname\relax
  \def\bibnamefont#1{#1}\fi
\expandafter\ifx\csname bibfnamefont\endcsname\relax
  \def\bibfnamefont#1{#1}\fi
\expandafter\ifx\csname citenamefont\endcsname\relax
  \def\citenamefont#1{#1}\fi
\expandafter\ifx\csname url\endcsname\relax
  \def\url#1{\texttt{#1}}\fi
\expandafter\ifx\csname urlprefix\endcsname\relax\def\urlprefix{URL }\fi
\providecommand{\bibinfo}[2]{#2}
\providecommand{\eprint}[2][]{\url{#2}}

\bibitem[{\citenamefont{Floquet}(1883)}]{floquet83}
\bibinfo{author}{\bibfnamefont{G.}~\bibnamefont{Floquet}},
  \bibinfo{journal}{Ann. Sci. \'{E}c. Norm. Sup\'{e}r.}
  \textbf{\bibinfo{volume}{12}}, \bibinfo{pages}{47} (\bibinfo{year}{1883}),
  \urlprefix\url{http://www.numdam.org/articles/10.24033/asens.220/}.

\bibitem[{\citenamefont{Shirley}(1965)}]{shirley65}
\bibinfo{author}{\bibfnamefont{J.~H.} \bibnamefont{Shirley}},
  \bibinfo{journal}{Phys. Rev.} \textbf{\bibinfo{volume}{138}},
  \bibinfo{pages}{B979} (\bibinfo{year}{1965}),
  \urlprefix\url{https://link.aps.org/doi/10.1103/PhysRev.138.B979}.

\bibitem[{\citenamefont{Ivanov et~al.}(2021)\citenamefont{Ivanov, Mote, Ernst,
  Equbal, and Madhu}}]{ivanov21}
\bibinfo{author}{\bibfnamefont{K.~L.} \bibnamefont{Ivanov}},
  \bibinfo{author}{\bibfnamefont{K.~R.} \bibnamefont{Mote}},
  \bibinfo{author}{\bibfnamefont{M.}~\bibnamefont{Ernst}},
  \bibinfo{author}{\bibfnamefont{A.}~\bibnamefont{Equbal}}, \bibnamefont{and}
  \bibinfo{author}{\bibfnamefont{P.~K.} \bibnamefont{Madhu}},
  \bibinfo{journal}{Progress in Nuclear Magnetic Resonance Spectroscopy}
  \textbf{\bibinfo{volume}{126-127}}, \bibinfo{pages}{17}
  (\bibinfo{year}{2021}), ISSN \bibinfo{issn}{0079-6565},
  \urlprefix\url{https://www.sciencedirect.com/science/article/pii/S0079656521000169}.

\bibitem[{\citenamefont{Tsuji et~al.}(2008)\citenamefont{Tsuji, Oka, and
  Aoki}}]{tsuji08}
\bibinfo{author}{\bibfnamefont{N.}~\bibnamefont{Tsuji}},
  \bibinfo{author}{\bibfnamefont{T.}~\bibnamefont{Oka}}, \bibnamefont{and}
  \bibinfo{author}{\bibfnamefont{H.}~\bibnamefont{Aoki}},
  \bibinfo{journal}{Phys. Rev. B} \textbf{\bibinfo{volume}{78}},
  \bibinfo{pages}{235124} (\bibinfo{year}{2008}),
  \urlprefix\url{https://link.aps.org/doi/10.1103/PhysRevB.78.235124}.

\end{thebibliography}

\end{document}


\title{Supplemental Material: \\ Decoding the drive-bath interplay: A guideline to enhance superconductivity}

\author{Rui Lin}
\affiliation{Institute for Theoretical Physics, ETH Z\"urich, 8093 Zurich, Switzerland}
\author{Aline Ramires}
\affiliation{Paul Scherrer Institut, 5232 Villigen PSI, Switzerland}
\author{R. Chitra}
\affiliation{Institute for Theoretical Physics, ETH Z\"urich, 8093 Zurich, Switzerland}

 
\maketitle

\section{The Shirley-Floquet structure for time-periodic systems}

The Shirley-Floquet formalism~\cite{floquet83,shirley65,ivanov21} can systematically solve the Schr\"odinger equation $[H(t) - i \partial_t] \Psi(t) = 0$ corresponding to a time-periodic one-body Hamiltonian
\begin{eqnarray}
	H(t) = \sum_{\ell\in\mathbb{Z}} H_{\ell} e^{i\ell\Omega t}.
\end{eqnarray}
It maps the operator $H-i\partial_t$ to a tight-binding model along the synthetic Floquet dimension as
\begin{eqnarray}
	\mathbb{H} = \sum_{\ell\in\mathbb{Z}} H_{\ell}\mathbb{F}_{\ell} - \Omega \mathbb{N},
\end{eqnarray}
via $e^{i\ell \Omega t}\mapsto(\mathbb{F}_\ell)_{m,n} = \delta_{m,n-\ell}$ and $i\partial_t \mapsto \mathbb{N}_{m,n} = i \delta_{m,n}$. Specifically, $\mathbb{F}_{\ell}$ and $\mathbb{N}$ are infinite-dimensional matrices in Floquet space. These matrices can be written down explicitly as
\begin{small}
	\begin{align}
		\begin{split}
			\mathbb{F}_0 = 
			\begin{pmatrix}
				& \vdots & \vdots & \vdots & \vdots & \vdots &  \\
				\cdots & 1 & 0 & 0 & 0 & 0 & \cdots \\
				\cdots & 0 & 1 & 0 & 0 & 0 & \cdots \\
				\cdots & 0 & 0 & 1 & 0 & 0 & \cdots \\
				\cdots & 0 & 0 & 0 & 1 & 0 & \cdots \\
				\cdots & 0 & 0 & 0 & 0 & 1 & \cdots \\
				& \vdots & \vdots & \vdots & \vdots & \vdots &  \\
			\end{pmatrix}, \quad
			\mathbb{F}_1 = 
			\begin{pmatrix}
				& \vdots & \vdots & \vdots & \vdots & \vdots &  \\
				\cdots & 0 & 1 & 0 & 0 & 0 & \cdots \\
				\cdots & 0 & 0 & 1 & 0 & 0 & \cdots \\
				\cdots & 0 & 0 & 0 & 1 & 0 & \cdots \\
				\cdots & 0 & 0 & 0 & 0 & 1 & \cdots \\
				\cdots & 0 & 0 & 0 & 0 & 0 & \cdots \\
				& \vdots & \vdots & \vdots & \vdots & \vdots &  \\
			\end{pmatrix}, \quad
			\mathbb{N} = 
			\begin{pmatrix}
				& \vdots & \vdots & \vdots & \vdots & \vdots &  \\
				\cdots & -2 & 0 & 0 & 0 & 0 & \cdots \\
				\cdots & 0 & -1 & 0 & 0 & 0 & \cdots \\
				\cdots & 0 & 0 & 0 & 0 & 0 & \cdots \\
				\cdots & 0 & 0 & 0 & 1 & 0 & \cdots \\
				\cdots & 0 & 0 & 0 & 0 & 2 & \cdots \\
				& \vdots & \vdots & \vdots & \vdots & \vdots &  \\
			\end{pmatrix},
		\end{split}
	\end{align}
\end{small}
and obey the following useful identities,
\begin{eqnarray}\label{eq:floquet_identities}
	\mathbb{F}_m\mathbb{F}_n = \mathbb{F}_{m+n},\quad 
	e^{\mathbb{F}_1} = \sum_{n}\frac{1}{n!}\mathbb{F}_n,\quad
	[\mathbb{F}_n,\mathbb{N}] = n\mathbb{F}_n,\quad 
	e^{\alpha\mathbb{F}_n}\mathbb{N}e^{-\alpha\mathbb{F}_n} = \mathbb{N}+\alpha n\mathbb{F}_n.
\end{eqnarray}
This formalism makes the time dependence of the problem implicit by providing an effective Hamiltonian $\mathbb{H}$, which can be solved using standard diagonalization techniques. The transformation operator $\mathbb{P}$ is composed of solely the matrices $\mathbb{F}_{\ell}$ in Floquet space, with an ansatz~\cite{ivanov21}
\begin{eqnarray}\label{eq:transformation_matrix}
	\mathbb{P} = \exp\left(\sum_{\ell\in\mathbb{Z}} S_{\ell} \mathbb{F}_{\ell}\right),
\end{eqnarray}
where $S_{\ell}$ are matrices in the space of $H_{\ell}$. Upon the diagonalization in Floquet space, we obtain the effective stroboscopic Hamiltonian.

The Green's function in the Shirley-Floquet formalism~\cite{tsuji08} can be obtained by inverting the Shirley-Floquet Hamiltonian,
\begin{eqnarray}
	\mathbb{G}(\omega) = [\omega \mathbb{F}_0 - \mathbb{H}]^{-1}.
\end{eqnarray}
The components of the Shirley-Floquet Green's function are related to the Green's function $G(t_1,t_2)$ in the Schr\"odinger representation as
\begin{eqnarray}
	G(\omega,\bar{t}) = \sum_{\ell\in\mathbb{Z}} G_{\ell}(\omega) e^{i\ell \Omega \bar{t}},
\end{eqnarray}
where $\bar{t}\equiv (t_1+t_2)/2$ is the time average and $\omega$ denotes the frequency associated to the time difference $\delta t \equiv t_2-t_1$.

Notably, the Shirley-Floquet formalism duplicates the time-dependent system into physically equivalent replicas. Due to this redundancy of the formalism, the physical quantities in the Shirley-Floquet formalism can all be written succinctly in the Wigner representation, see Eq.~(4) in the main text. For example, the Shirley-Floquet Green's function $\mathbb{G}(\omega)$ can be written in terms of its Wigner representation $G_{\ell}(\omega)$ as
\begin{eqnarray}\label{eq:full_G_matrix}
	\mathbb{G}(\omega) = 
	\begin{pmatrix}
		& \vdots & \vdots & \vdots & \vdots & \vdots &  \\
		\cdots & G_0(\omega+2\Omega) & G_1(\omega+3\Omega/2) & G_2(\omega+\Omega) & G_3(\omega+\Omega/2) & G_4(\omega) & \cdots \\
		\cdots & G_{-1}(\omega+3\Omega/2) & G_0(\omega+\Omega)  & G_1(\omega+\Omega/2) & G_2(\omega) & G_3(\omega-\Omega/2) & \cdots \\
		\cdots & G_{-2}(\omega+\Omega) & G_{-1}(\omega+\Omega/2) & G_0(\omega)  & G_1(\omega-\Omega/2) & G_2(\omega-\Omega) & \cdots \\
		\cdots & G_{-3}(\omega+\Omega/2) & G_{-2}(\omega) & G_{-1}(\omega-\Omega/2) & G_0(\omega-\Omega)  & G_1(\omega-3\Omega/2) & \cdots \\
		\cdots & G_{-4}(\omega) & G_{-3}(\omega-\Omega/2) & G_{-2}(\omega-\Omega) & G_{-1}(\omega-3\Omega/2) & G_0(\omega-2\Omega)  & \cdots \\
		& \vdots & \vdots & \vdots & \vdots & \vdots &  \\
	\end{pmatrix}.
\end{eqnarray}

We conclude this section by proving a crucial statement in the main text, $\mathbb{P}^\dagger \bbrho(\omega) \mathbb{P} \neq \bbrho(\omega)$. For this purpose, we rewrite the transformation operator in the basis of $\mathbb{F}_\ell$ as $\mathbb{P} = \sum_{\ell\in\mathbb{Z}} P_{\ell} \mathbb{F}_\ell$, see Eq.~\eqref{eq:transformation_matrix}, where $P_{\ell}$ are matrices in the same space as $H_{\ell}$. In this basis, we can observe a relation, $\bbrho(\omega)\mathbb{F}_{\ell} = \mathbb{F}_{\ell} \bbrho(\omega+\ell\Omega)$, between the thermal distribution $\rho(\omega)$ and the basis matrices $\mathbb{F}_{\ell}$. This relation allows us to write down
\begin{eqnarray}
	\mathbb{P}^\dagger \bbrho(\omega) \mathbb{P} = \sum_{m,n\in\mathbb{Z}} P^\dagger_m P_n \mathbb{F}_{n-m}  \bbrho(\omega+n\Omega),
\end{eqnarray}
where we have further used the fact that $[P_\ell,\bbrho]=0$ because they are in different spaces. When $\mathbb{P}$ indeed describes a time-periodic frame transformation, there exists $n\neq0$ such that $P_{n}\neq 0$. In this case, it can be seen that $\mathbb{P}^\dagger \bbrho(\omega) \mathbb{P} \neq \bbrho(\omega)$. For example, in the simplest case where $P_n = \delta_{n,1}$, we obtain $\mathbb{P}^\dagger \bbrho(\omega) \mathbb{P} = \bbrho(\omega+\Omega)$. This indicates that in any time-dependent rotating reference frame, the thermal bath no longer provides an equilibrium thermal distribution.

\section{Retarded Green's function for commuting drive}
\label{sec:SC_green}

We use the Shirley-Floquet formalism developed above to calculate the Green's function corresponding to the Hamiltonian 
\begin{eqnarray}
	H^{\tau} = \epsilon \tau_z + \Delta_0 \tau_y + a\cos(\Omega t)\tau_0,
\end{eqnarray}
where we have suppressed the spin space for the current discussion, because it does not affect the commutation relation.
We provide an analytically tractable example for many of the driven-dissipative effects discussed in the main text, particularly for the results related to a commuting drive, $\zeta=1$.
Note that, because the drive is in the normal state, i.e. $\tau_0$ or $\tau_z$ Nambu sector, it commute with the dispersion $\tau_z$ by construction. Therefore, our discussions below are valid for all drives which commute with the gap operator.

In the Shirley-Floquet representation, the Hamiltonian of the driven system can be rewritten as
\begin{eqnarray}
	\mathbb{H} = (\epsilon\tau_z+\Delta_0\tau_y)\mathbb{F}_0 + \frac{a}{2}\tau_0(\mathbb{F}_1+\mathbb{F}_{-1}) - \Omega \mathbb{N}.
\end{eqnarray}
The commutativity between the drive and the rest of the Hamiltonian allows us to diagonalize the Hamiltonian using
\begin{align}
		\mathbb{P} = \exp\left[-\frac{a}{2\Omega}\tau_0(\mathbb{F}_1-\mathbb{F}_{-1})\right] 
		=\sum_{m=0}^{\infty}\sum_{n=0}^{\infty}\frac{(-a/2\Omega)^{m}}{m!}\frac{(a/2\Omega)^{n}}{n!} \tau_0\mathbb{F}_m \mathbb{F}_{-n}
		= \sum_{n=-\infty}^{\infty} J_{n}\left(-\frac{a}{\Omega}\right)\tau_0\mathbb{F}_n
\end{align}
as
\begin{eqnarray}
	\mathbb{H}' = \mathbb{P}^\dagger \mathbb{H} \mathbb{P}= (\epsilon\tau_z+\Delta_0\tau_y)\mathbb{F}_0 - \Omega_0 \mathbb{N},
\end{eqnarray}
where $J_n$ are the Bessel functions of the first kind. Particularly, we have used the identities in Eq.~\eqref{eq:floquet_identities} and the Taylor expansion $J_{n\ge0}(2x) = \sum_{k=0}^{\infty} \frac{(-1)^k}{k!(n+k)!}x^{n+2k}$. It can be subsequently recognized that the Hamiltonian in the rotating frame is exactly the same as the Hamiltonian of the non-driven system $\mathbb{H}' = \tilde{\mathbb{H}} \equiv \mathbb{H}(a=0)$.
Since $\mathbb{H}'$ is diagonal in Floquet space, its Green's function can be solved by inverting the individual static replicas as [cf.\ Eq.~(14) of the main text]
\begin{eqnarray}\label{eq:append_SC_green_G_static}
	G'^{R}_{\ell}(\omega) = \tilde{G}^{R}_{\ell}(\omega) = \delta_{\ell,0}\tilde{G}^{R}(\omega) = \frac{\omega_+ \tau_0 + \epsilon \tau_z + \Delta_0 \tau_y}{\omega_+^2-E^2} \delta_{\ell,0},
\end{eqnarray}
with $E=\sqrt{\epsilon^2+\Delta_0^2}$. 
The Green's function of the driven system in the lab frame 
\begin{eqnarray}
	\mathbb{G}^{R} = \mathbb{P} \mathbb{G}'^{R} \mathbb{P}^\dagger,
\end{eqnarray}
can finally be solved as
\begin{eqnarray}
	G^{R}_{\ell}(\omega) = \sum_{\tilde{k}\in\mathbb{Z}}(-)^{\tilde{k}} \tilde{G}^{R}\left(\omega+\frac{2\tilde{k}-k}{2}\Omega\right)J_{\tilde{k}}(a/\Omega)J_{k-\tilde{k}}(a/\Omega).
\end{eqnarray}
This is equivalent to the expression shown in the main text, and is also consistent with the approximations Eq.~(19) in the main text, as can be shown using the Taylor expansion of $J_n(x)$.

In the absence of the coupling to a thermal bath, the different Shirley-Floquet Hamiltonians describing the same system in different frames, i.e.,  $\mathbb{H}$ and $\mathbb{H}'$, should be physically equivalent. This is because they are related by a transformation operator $\mathbb{P}$. This can indeed be reflected in the retarded/advanced Green's functions and thus the spectral functions, when all excitations are taken into account and integrated over,
\begin{eqnarray}
	\int_{-\infty}^{\infty} d\omega (G^{R})_{\ell}(\omega) 
	=\int_{-\infty}^{\infty} d\omega (G'^{R})_{\ell}(\omega), \quad \forall \ell,
\end{eqnarray}
where we have used the identity $\sum_{n=-\infty}^{\infty} J_{n}(x) J_{n+\ell}(x) = \delta_{0,\ell}$. Specifically for the commuting case with $\zeta=1$, because $\mathbb{H}'=\mathbb{H}(a=0)$, the drive has no physical effect and can be rotated out completely in the absence of the thermal bath, as soon as the system reaches equilibrium.

Nevertheless, when the coupling to the thermal bath is considered explicitly, we need to refer to the Keldysh Green's function and the response function. These quantities are indeed sensitive to the drive, 
\begin{eqnarray}
	\int_{-\infty}^{\infty} d\omega \mathcal{R}^\mathrm{an}(\omega) 
	<\int_{-\infty}^{\infty} d\omega \tilde{\mathcal{R}}^\mathrm{an}(\omega),\quad a>0
\end{eqnarray}
where we have used the concavity of $\tanh(x),\, \forall x>0$. This indicates that a commuting drive generally suppresses the excitaitons and effectively heats up the system.

\section{Oscillating components of the order parameter}
\label{sec:append_SC_high_order}

We discuss the behaviors of the oscillating components of the order parameter $\Delta_{\ell\neq0}$. For clarity of the discussion, we can rewrite the gap equation [Eq.~(9) of the main text] for each Fourier components of the gap as 
\begin{eqnarray}\label{eq:gap_equation_components}
	\Delta_{\ell} = ig\int d\omega \sum_{\mathbf{k}} \tr_{\tau\sigma F} \left(\hat{\Delta}_{\mathbf{k}} \mathbb{G}^{K}_{\mathbf{k}}\mathbb{F}_{-\ell}\right)
\end{eqnarray}
Notably, the Floquet space has a repeating structure as expressed by the Wigner representation, cf.\ Eq.~\eqref{eq:full_G_matrix}, particularly upon integration over $\omega$. Therefore, the trace in Floquet space $\tr_F \mathbb{A}$ essentially extracts the entry $\mathbb{A}_{(0,0)}$ of the Floquet matrix.

\subsection{Frame invariance of oscillation}

We start by proving the frame invariance of this gap equation, with a general change of time-dependent reference frame expressed by its operator $\mathbb{P}$ in Floquet space. Upon the transformation, all the normal state, the superconducting gap, and the bath are subject to the mapping,
\begin{eqnarray}
	\hat{\Delta}_{\mathbf{k}} \mathbb{G}^{K}_{\mathbf{k}} \mapsto \mathbb{P}^\dagger\hat{\Delta}_{\mathbf{k}} \mathbb{G}^{K}_{\mathbf{k}} \mathbb{P}.
\end{eqnarray}
However, because generally $\mathbb{P}$ can be expanded in the basis of $\mathbb{F}_\ell$ [cf.\ Eq.~\eqref{eq:transformation_matrix}] and thus $[\mathbb{P},\mathbb{F}_\ell]=0$, we immediately observe
\begin{eqnarray}
	\Delta_{\ell} \mapsto ig\int d\omega \sum_{\mathbf{k}} \tr_{\tau\sigma F} \left(\mathbb{P}^\dagger\hat{\Delta}_{\mathbf{k}} \mathbb{G}^{K}_{\mathbf{k}}\mathbb{F}_{-\ell}\mathbb{P}\right) = \Delta_{\ell}
\end{eqnarray}
upon the change of reference frame. Particularly, we have used the invariance of trace upon similarity transformation in Floquet space.

\subsection{Dominance of the time-constant component of the gap}

\begin{figure}[t]
	\centering
	\begin{tikzpicture}[scale=1]		
		
		\draw[line width=0.6mm,->] (-3,0) -- (3,0);
		\draw[line width=0.6mm,->] (0,-1.5) -- (0,1.5);
		\draw[line width=0.4mm,dashed] (-3,1) -- (3,1);
		\draw[line width=0.4mm,dashed] (-3,-1) -- (3,-1);
		\draw[domain=-3:3, smooth, variable=\x, red, line width=0.6mm] plot({\x}, {tanh(3*\x)});
		\draw[domain=-3:0, smooth, variable=\x, blue, line width=0.6mm] plot({\x}, {(tanh((\x+1)*3)-1)/2});
		\draw[domain=0:3, smooth, variable=\x, blue, line width=0.6mm] plot({\x}, {(tanh((\x-1)*3)+1)/2});
		\draw[domain=-3:0, smooth, variable=\x, violet, line width=0.6mm] plot({\x}, {(tanh((\x+2)*3)-1)/2});
		\draw[domain=0:3, smooth, variable=\x, violet, line width=0.6mm] plot({\x}, {(tanh((\x-2)*3)+1)/2});
		
		\draw[line width=0.4mm,dashed] (1,0) -- (1,1);
		\draw[line width=0.4mm,dashed] (2,0) -- (2,1);
		\draw[line width=0.4mm,dashed] (-1,0) -- (-1,-1);
		\draw[line width=0.4mm,dashed] (-2,0) -- (-2,-1);
		
		\node[] at (3.2,0) {$\omega$};
		\node[] at (0,1.7) {$T_{\ell}$};
		\node[] at (0.2,-0.2) {$0$};
		\node[] at (1,-0.2) {$\Omega/2$};
		\node[] at (2,-0.2) {$\Omega$};
		\node[] at (-1,0.2) {$-\Omega/2$};
		\node[] at (-2,0.2) {$-\Omega$};
		\node[] at (-0.2,0.8) {$1$};
		\node[] at (-0.3,-1.2) {$-1$};
	\end{tikzpicture}
	\caption{
		The functions $T_{\ell}(\omega)$ at temperature $T=\Omega/3$, with (red) $\ell=0$, (blue) $\ell=1$, and (purple) $\ell=2$.
	}
	\label{fig:append_SC_Tl}
\end{figure}
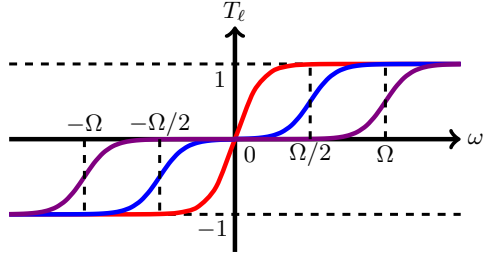

We now proceed to prove that the time-constant component of the gap $\Delta_0$ is indeed dominating over the oscillating components $\Delta_{\ell\neq0}$. For this purpose, we further rewrite the gap equations Eq.~\eqref{eq:gap_equation_components} in the specific choice of the lab frame and $\hat{\Delta}_\mathbf{k} = d_{\mathrm{k}}\tau_y\sigma_y$,
\begin{align}
	\Delta_{\ell} = ig \int_{-\Omega_C}^{\Omega_C} d\omega T_{\ell}\left(\frac{\omega}{2T},\Omega\right)\sum_{\mathbf{k}} d_{\mathrm{k}} \tr_{\tau\sigma}[(G_{\mathbf{k}}^{R})_{\ell}\tau_y\sigma_y],
\end{align}
where
\begin{align}
	T_{\ell}\left(\frac{\omega}{2T},\Omega\right) = \frac{1}{2}\tanh\left(\frac{\omega-\ell\Omega/2}{2T}\right)+\frac{1}{2}\tanh\left(\frac{\omega+\ell\Omega/2}{2T}\right),
\end{align}
Specifically, we have also taken explicitly into account the generalized fluctuation-dissipation theorem [Eq.~(5) of the main text], and the cutoff frequency $\Omega_C$, and formulate the equation with the retarded Green's function $G^R$, instead of the Keldysh Green's function $G^K$.

For clarity, we consider explicitly the choice of the Hamiltonian Eq.~(11) in the main text,
\begin{eqnarray}
	H_{0,\mathbf{k}} = \epsilon_{\mathbf{k}} \tau_z\sigma_0,\quad  H_{1} = H_{-1} = \frac{a_\mathbf{k}}{2} \tau_\mu\sigma_\nu, \quad H_{|\ell|\ge2}^{\tau\sigma}=0,
\end{eqnarray}
where $\mu\in\{0,z\}$ and $\nu\in\{0,x,y,z\}$.
We also focus our discussion at low temperatures $T\ll\Omega$, and with the drive at near resonant to the cutoff frequency $\Omega_C\approx\Omega$.

The behaviors of $T_{\ell}$ are depicted in Fig.~\ref{fig:append_SC_Tl}. At zero temperature, it behaves like a sign function $\sgn(\omega)$, except being zero $T_{\ell}(\omega)=0$ in the interval $\omega\in(-\ell\Omega/2,\ell\Omega/2)$. This indicates that the excitations of $(G_{\mathbf{k}}^{R})_{\ell}$ within this interval do not contribute to the order parameter $\Delta_\ell$. Considering also the fact that the driving frequency is chosen to be in close resonance with the cutoff frequency $\Omega\approx\Omega_C$, we immediately obtain $\Delta_{|\ell|\ge3}=0$ at zero temperature. At finite temperature, $T_{\ell}$ become smoothened in a similar way as the Fermi-Dirac distribution, activating an exponentially tiny contribution in $\Delta_{|\ell|\ge3}\sim \exp\left[-\frac{(|\ell|/2-1)\Omega}{T}\right]$. These components are thus negligible in comparison to the lower harmonics $\Delta_{\pm1,\pm2}$ in the low temperature regime.

We first consider $\Delta_{\pm1}$, which are related to the Green's functions in the first Wigner families $(G_{\mathbf{k}}^{R})_{\ell=\pm1}$. To the order of $O(a^2/\Omega^2)$, these Green's functions are approximated by [cf.\ Eq.~(13) of the main text]
$(G_{\mathbf{k}}^{R})_{\ell=\pm1}(\omega) \approx \frac{a_\mathbf{k}}{2}\tilde{G}_{\mathbf{k}}^{R}(\omega\mp\Omega/2) \tau_\mu\sigma_\nu\tilde{G}_{\mathbf{k}}^{R}(\omega\pm\Omega/2)$.
Interestingly, $\Delta_{\pm1}$ vanishes unless the drive is in $\tau_0\sigma_0$ sector or the same sector as the dispersion, i.e. $\tau_z\sigma_0$.
Otherwise, $\tr_{\tau\sigma}[(G_{\mathbf{k}}^{R})_{\ell}\tau_y\sigma_y]=0$ as the drive rotates the order parameter into a sector orthogonal to the interaction channel. This argument in fact also applies to all odd Fourier components of the gap, $\Delta_{\ell},\,\ell\in2\mathbb{Z}+1$.  As a result of the vanishing $\Delta_{\pm1}$, a frequency doubling in the order parameter is predicted.

\begin{figure}[t]
	\centering
	\includegraphics[width=0.3\columnwidth]{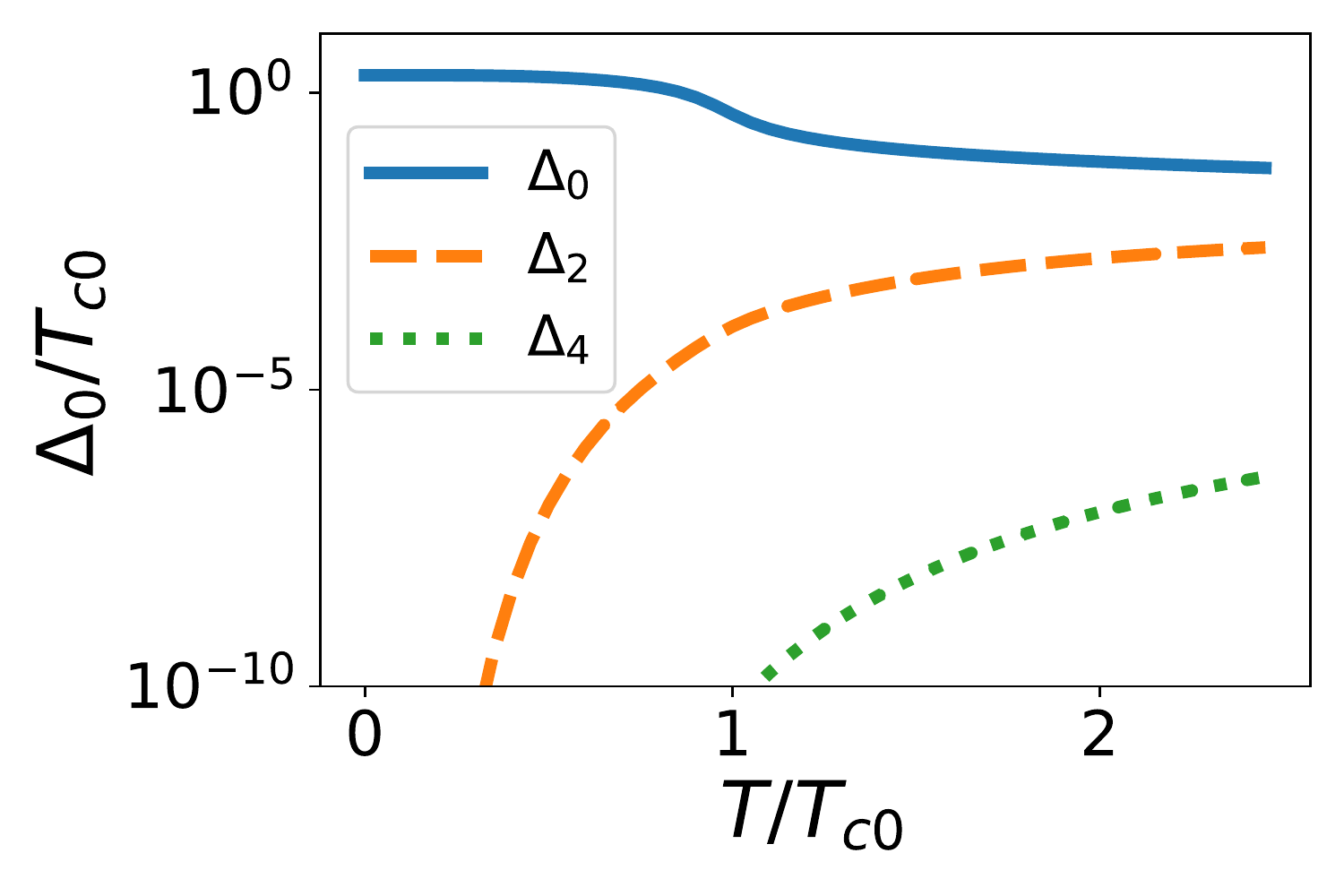}
	\caption{
		The oscillating components $\Delta_2$ and $\Delta_4$ as functions of temperature. The results are obtained by numerically solving the full gap equation [Eq.~(9) of the main text] for a flat-band superconductor $\epsilon_{\mathbf{k}}=0$, subject to an anti-commuting drive $\zeta=-1$ with amplitude $a=0.3\Omega$. The result for the time-averaged component $\Delta_0$ is reproduced from Fig.~1(b) of the main text, and $\Delta_1=\Delta_3=0$. The oscillating components are indeed negligible compared to the time-averaged component.
	}
	\label{fig:Delta_oscillate}
\end{figure}

We then consider $\Delta_{\pm2}$ related to $(G_{\mathbf{k}}^{R})_{\ell=\pm2}$. As a result of the cutoff $\Omega_C$, the dominant excitations of the Green's function are the ones located at $\omega=\pm(\Omega-E_{\mathbf{k}})$. A similar argument as above shows that $\Delta_{\pm2}$ generally do not vanish, but rather scale as $a^2/\Omega^2$. Considering the thermal distribution which behaves as $T_{\ell}(\omega)\approx \frac{1}{2}[1-\tanh(E_{\mathbf{k}}/2T)]$ at $\omega=\pm(\Omega-E_{\mathbf{k}})$, we obtain the asymptotic behavior
\begin{eqnarray}
	\Delta_{\pm2} \sim \frac{a^2}{\Omega^2}\left[1-\tanh\left(\frac{E_{\mathbf{k}}}{2T}\right)\right].
\end{eqnarray}
At very low temperature $T\ll E_{\mathbf{k}}$, $\Delta_{\pm2}$ are thus suppressed exponentially in a similar way as all other oscillating components, regardless of the driving amplitude $a/\Omega$, and our a priori assumption that the constant component $\Delta_0$ dominates is thus valid.
On the other hand, when the temperature is comparable to the energy scale of the system $T\sim E_{\mathbf{k}}\ll \Omega$, $\Delta_{\pm2}\sim a^2/\Omega^2$ become sizable, and its omission might be inappropriate. This is confirmed in our numerical solution of the gap equation for the driven-dissipative flat-band superconductors. See Fig.~\ref{fig:Delta_oscillate}, which depicts $\Delta_2$ and $\Delta_4$ as functions of temperature.
Nevertheless, this deviation should only be quantitative, and does not invalidate our arguments.

We have thus shown the insignificant contributions of the higher harmonics. This also indicates the robustness of our scheme with respect to the existence of higher harmonics in the driving. 
Even though these higher harmonics are unavoidable in realistic experimental implementations, they only induce shifted poles beyond the third temporal Brillouin zones, which are suppressed by the cutoff frequency.

\section{Dispersive superconductor in the weak coupling limit}

We complement our results for flat-band superconductors in the main text by more general calculations for dispersive superconductors. More specifically, we consider the weak coupling limit where the summation over momentum can be replaced by an integral over energy as 
$\sum_{\mathbf{k}} \to N(\epsilon_F) \int_{-\Omega_C}^{\Omega_C} d\epsilon$, where $N(\epsilon_F)$ is the density of states near the Fermi surface.
Similar to Eq.~(21) of the main text, using the Dyson equation [Eq.~(13) of the main text], we evaluate the anomalous spectral and response functions up to the order of $O(a^2/\Omega^2)$ for the case of an anti-commuting drive,
\begin{subequations}
	\begin{align}
		\mathcal{R}^{\mathrm{an}}_{\epsilon}(\omega) =& \tanh\left(\frac{\omega}{2T}\right)\mathcal{A}^{\mathrm{an}}_{\epsilon}(\omega) \\
		\mathcal{A}^{\mathrm{an}}_{\epsilon}(\omega) \approx& \left(\frac{\Delta_0}{E}-\frac{a^2\Delta_0}{\Omega^2}B_0\right)[L_\Sigma(\omega-E)-L_\Sigma(\omega+E)] \\
		& -\frac{a^2\Delta_0}{\Omega^2}B_+[L_\Sigma(\omega-\Omega-E) - L_\Sigma(\omega+\Omega+E)] \nonumber\\
		& +\frac{a^2\Delta_0}{\Omega^2}B_-[L_\Sigma(\omega-\Omega+E) - L_\Sigma(\omega+\Omega-E)] \nonumber
	\end{align}
with
	\begin{align}
		B_0 =& \frac{(16\epsilon^2E^4+4(\Delta_0^2-4\epsilon^2)E^2\Omega^2)+(3\epsilon^2+\Delta_0^2)\Omega^4}{2E^3(\Omega^2-4E^2)^2} \\
		B_{\pm} =& \frac{\Omega^2-4\epsilon^2}{4 E(2E\pm\Omega)^2},
	\end{align}
\end{subequations}
where we have kept $\epsilon$ finite, and $E=\sqrt{\epsilon^2+\Delta_0^2}$.

 \begin{figure}[t]
	\centering
	\includegraphics[width=0.5\columnwidth]{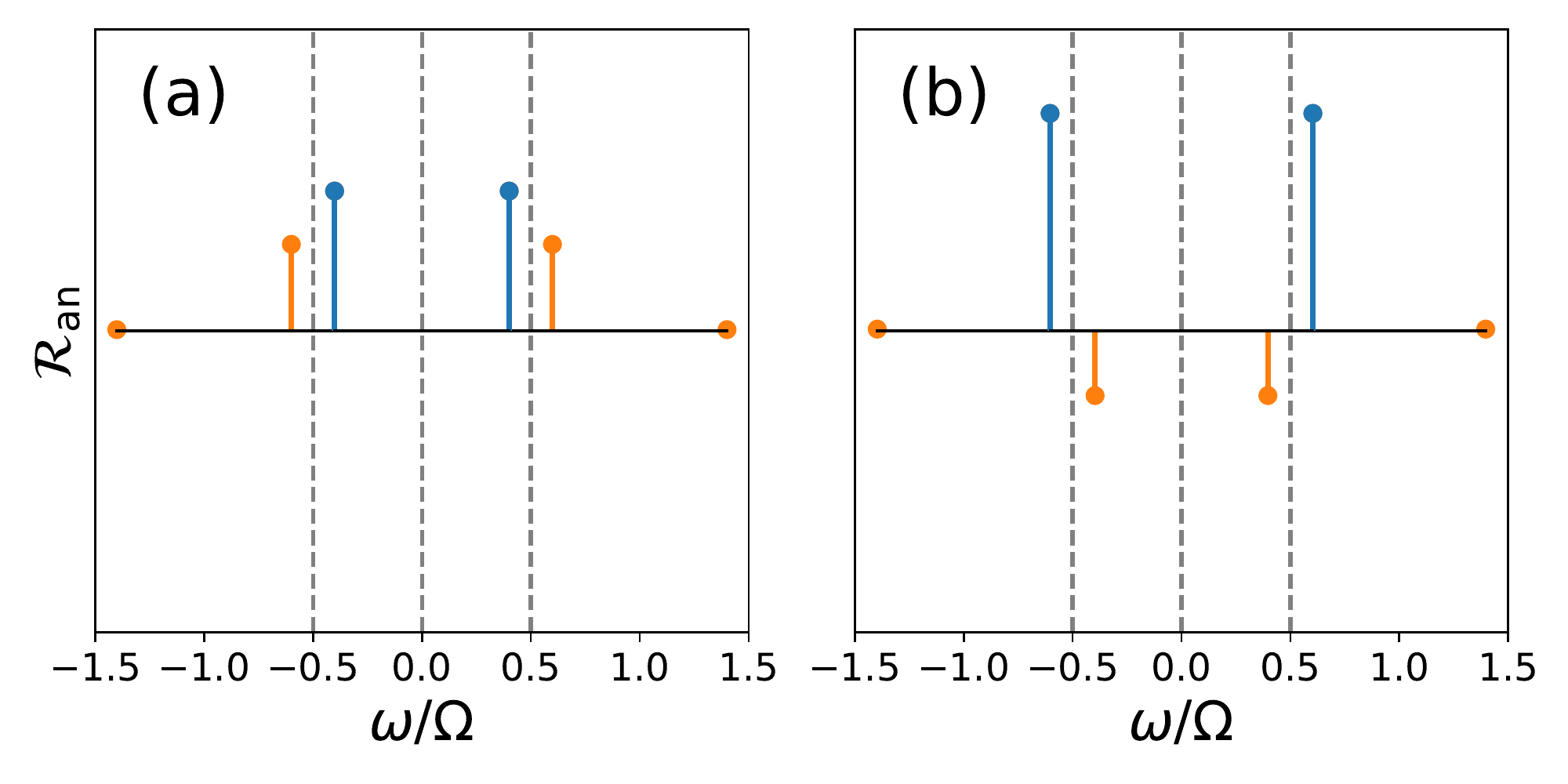}
	\caption{
		Anomalous response function of the driven-dissipative superconductor with an anti-commuting drive $\zeta=-1$. When the driving frequency is close to twice of the dispersion $\Omega\approx 2\epsilon$, the drive induces parametric resonance, which makes the weights of the shifted excitations also become significant. This can be observed when $\epsilon$ approaches $\Omega/2$ from below and above, i.e., for both (a) $\epsilon=0.4\Omega$ and (b) $\epsilon=0.6\Omega$. In comparison, the excitation shifted in the other direction is off resonance, and carries much insignificant weight.
		Here, the unshifted excitations are marked in blue, while the shifted excitations are marked in orange.
	}
	\label{fig:response_resonance}
\end{figure}

\begin{figure}[t]
	\centering
	\includegraphics[width=0.5\columnwidth]{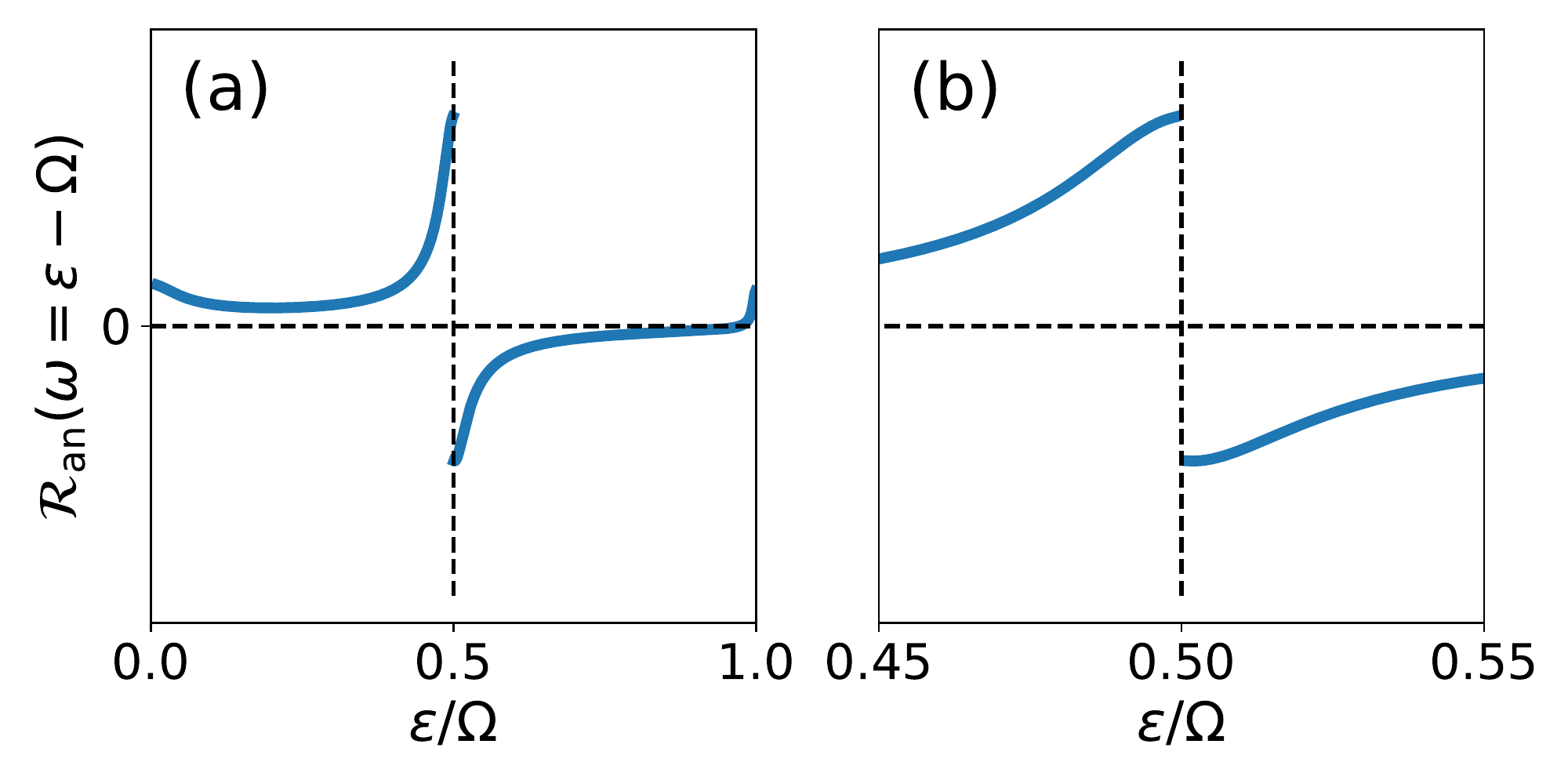}
	\caption{
		(a) The response function at the shifted particle excitation $\mathcal{R}_\mathrm{an}(\omega=\epsilon-\Omega)$ as a function of $\epsilon/\Omega$ when $\Delta_0\ll\Omega$ and $a\ll\Omega$ are fixed. In the vicinity of $\epsilon\approx\Omega/2$, a parametric resonance takes place between the shifted particle excitation and the unshifted hole excitation. This leads to a significant enhancement in the response function. (b) A zoom-in in the vicinity of the resonance regime. The enhancement is indeed not diverging.
	}
	\label{fig:response_dispersion}
\end{figure}

We comment on another qualitative distinction between commuting ($\zeta=1$) and anti-commuting ($\zeta=-1$) drives. With the anti-commuting drive, a parametric resonance occurs when the driving frequency is twice of the dispersion, $\Omega=2E$. In other words, the driving frequency exactly matches the energy difference between the particle and hole excitations. Without loss of generality, we discuss this parametric resonance based on the particle excitation at $E$.
As discussed in the main text and illustrated in Fig.~\ref{fig:response_resonance}, the particle excitation appears as a hole excitation after its energy is shifted by $\Omega$ by the anti-commuting drive. This shifted hole excitation at $E-\Omega$ now becomes almost degenerate with the unshifted hole excitation at $-E$, leading them resonant with each other. 
As a result of such resonance, the weight of the shifted hole excitation in the response function becomes comparable to the weights of the unshifted ones. In comparison, the hole excitation at $E+\Omega$, which is off resonant with other excitations, has insignificant weight. This parametric resonance can be observed when $\epsilon$ is slightly above or below $\Omega/2$. 

The diverging parametric resonance seen in the low-order Dyson equation becomes smoothened when all orders are taken into account. We show the numerically evaluated weight of the shifted excitation at $\mathcal{R}^{\mathrm{an}}_{\epsilon}(\epsilon-\Omega)$ as a function of the dispersion $\epsilon$ in Fig.~\ref{fig:response_dispersion}. The resonance at $\epsilon=\Omega/2$ leads to a significant enhancement, which does not diverge [Fig.~\ref{fig:response_dispersion}(b)].
Moreover, a sign change in $\mathcal{R}^\mathrm{an}$ is seen as $\epsilon$ goes from below $\Omega/2$ to above. When all excitations are taken into account and integrated over, the enhanced contributions cancel each other out. The parametric resonance induced by the anti-commuting drive thus has no significant consequences in our scheme.

\begin{figure}[t]
	\centering
	\includegraphics[width=0.4\columnwidth]{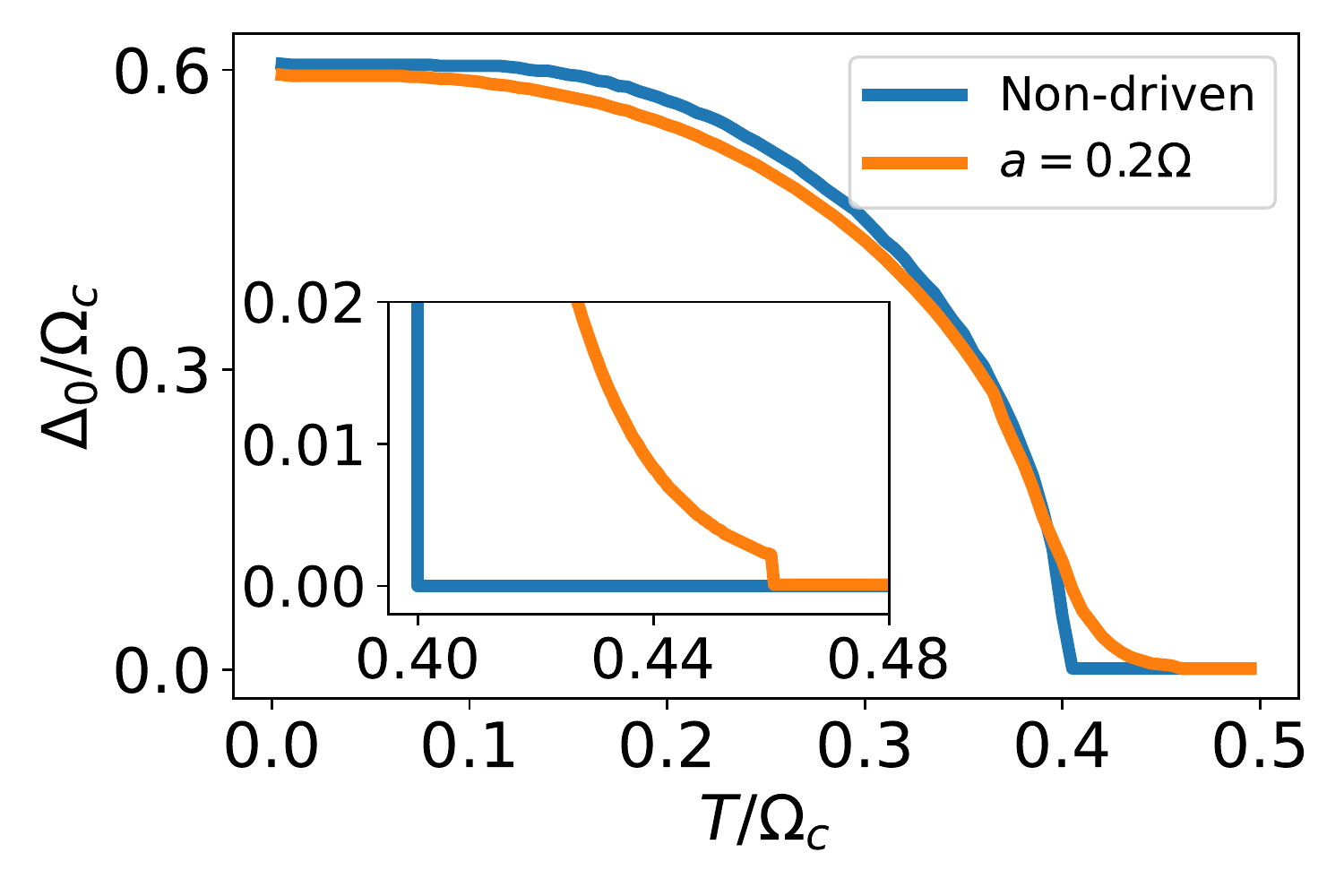}
	\caption{
		The time-averaged gap $\Delta_0$ as a function of temperature for a dispersive superconductor, subject to an anti-commuting drive $\zeta=-1$ with amplitude $a=0.2\Omega$. The gap of a non-driven superconductor is also shown for comparison.
		A zoom-in in the vicinity of the transition temperature is shown in the inset, where a first-order jump is clearly seen for the driven case.
	}
	\label{fig:Delta_dispersive}
\end{figure}

Therefore, we can proceed to investigate the superconducting transition temperature and the validity of its enhancement for dispersive systems
For this purpose, we write down the self-consistency gap equation corresponding to the response function, 
\begin{subequations}
	\begin{align}
		\frac{1}{gN(\epsilon_F)} =& I_1 + \frac{a^2}{\Omega^2}I_2 \\
		I_1 =& \int_0^{\Omega-\delta\Omega} d\epsilon \frac{1}{E}\tanh\left(\frac{E}{2T_c}\right)\\
		I_2 =& \int_{0}^{2\Omega+\delta\Omega} d\epsilon B_- \tanh\left(\frac{\Omega-E}{2T_c}\right) - \int_{0}^{\delta\Omega} d\epsilon B_+ \tanh\left(\frac{\Omega+E}{2T_c}\right) 
		- \int_{0}^{\Omega-\delta\Omega} d\epsilon B_1 \tanh\left(\frac{E}{2T_c}\right).
	\end{align}
\end{subequations}
The transition temperature of the driven superconductor can be estimated by imposing $\Delta_0=0$, which yields the equation for $T_c$ as
\begin{subequations}
	\begin{eqnarray}
		I_1(\Delta_0=0) &\approx& \ln\left(\frac{\Omega}{2T_c}\right)+\ln\left(\frac{4}{\pi}\right)+\gamma \\
		I_2(\Delta_0=0)  &\approx& \frac{1}{4} \left[\ln \left(\frac{\Omega}{\delta \Omega}\right) - 6\ln\left(\frac{\Omega}{T_c}\right) + \ln\left(\frac{3^5 \pi^6}{2^{11}}\right) - 6\gamma\right],
	\end{eqnarray}
\end{subequations}
where $\gamma\approx0.577$ is Euler's constant.
The transition temperature $T_c$ for the driven superconductor is thus approximately related to the non-driven critical temperature $T_{c0}=T_c(a=0)$ as
\begin{eqnarray}
	\frac{T_c}{T_{c0}} \approx \left(\frac{3.573 T_{c0}^6}{\Omega^5\delta\Omega}\right)^{\frac{a^2}{4\Omega^2}}.
\end{eqnarray}
We have thereby shown that the driven-dissipative enhancement is qualitatively recovered for the dispersive superconductor. This is again confirmed by a numerical solution of the gap equation, as shown in Fig.~\ref{fig:Delta_dispersive}.
In comparison to the flat band superconductor [Eq.~(23) of the main text], the dispersive superconductor is subject to a much weaker enhancement. A much more resonantly tuned driving is required for achieving the same enhancement in $T_c$. This is essentially because our scheme works most significantly for states lying in the vicinity of the Fermi surface, whose density is reduced for a dispersive superconductor.

\bibliographystyle{apsrev}
\bibliography{References}